\documentclass[letterpaper,english,reprint, aps]{revtex4-2}
\pdfoutput=1
\usepackage[latin9]{inputenc}
\usepackage{textcomp}
\usepackage{amsmath}
\usepackage{amsthm}
\usepackage{amssymb}
\usepackage{graphicx}
\usepackage{subscript}
\usepackage{xcolor}
\usepackage{soul}
\usepackage{multirow}

\makeatletter

\pdfpageheight\paperheight
\pdfpagewidth\paperwidth

\numberwithin{equation}{section}
\numberwithin{figure}{section}

\makeatother

\usepackage{babel}
\begin{document}
\title{Two limit cases of twisted hBN bilayers and their excitonic response}
\author{J. C. G. Henriques$^{1}$, B. Amorim$^{1}$, R. M. Ribeiro$^{1,2}$
and N. M. R. Peres$^{1,2}$}
\affiliation{$^{1}$Department and Centre of Physics, University of Minho, Campus
of Gualtar, 4710-057, Braga, Portugal}
\affiliation{$^{2}$International Iberian Nanotechnology Laboratory (INL), Av. Mestre
Jose Veiga, 4715-330, Braga, Portugal}
\begin{abstract}
In this paper we discuss the optical response due to the excitonic
effect of two types of hBN bilayers: AB and AA'. Understanding the
properties of these bilayers is of great utility to the study of twisted
bilayers at arbitrary angles, since these two configurations correspond
to the limit cases of $0^\circ$ and $60^\circ$ rotation. To obtain the excitonic
response we present a method to solve a four-band Bethe-Salpeter equation, by casting it
into a 1D problem, thus greatly reducing the numerical burden of the
calculation when compared with strictly 2D methods. We find results in good agreement with \emph{ab initio} calculations already
published in the literature for the AA' bilayer, and predict the excitonic conductivity of the AB bilayer, which remains largely unstudied. The main difference
in the conductivity of these two types of bilayers is the appearance
of a small, yet well resolved, resonance between two larger ones in
the AB configuration. This resonance is due to a mainly interlayer
exciton, and is absent in the AA' bilayer. Also, the conductivity
of the AB bilayer is due to both intralayer and interlayer excitons and is dominated by p-states, while intralayer s-states
are the relevant ones for the AA' configuration, like in a monolayer.
The effect of introducing a bias in the AA' bilayer is also discussed.
\end{abstract}
\maketitle

\section{Introduction}

In its monolayer form, hexagonal boron nitride (hBN) is an insulator
with a direct band gap located at the vertices of the first Brillouin
zone, with a magnitude close to 6 eV \citep{caldwell2019photonics}.
Contrarily to transition metal dichalcogenides (TMDs) \citep{Wang2018Colloquium},
the lack of heavy metals leads to a rather small spin orbit coupling
effect. The simplicity of its band structure and the large band gap,
make this an excellent material for the exploration of fundamental
physics. Due to its structural similarity with graphene, hBN monolayers
are often used as a substrate for graphene \citep{dean2010boron,kretinin2014electronic,ashhadi2017electronic},
or to encapsulate other materials, protecting them from the environment
\citep{epstein2020highly}. On their own, hBN monolayers are mostly
studied because of their optical response dominated by excitonic resonances.
In the simplest possible picture, an exciton is formed when an electron
is promoted to the conduction band, leaving a hole in the valence
band. These two particles, having opposite charges, interact via an
electrostatic potential \citep{Cudazzo2011}, leading to the formation
of a bound state. This composite quasi-particle is then responsible
for the optical absorption inside the band gap of the material. This
optical response has been essential for the exploration of hBN in
deep-UV optoelectronics \citep{watanabe2004direct,kubota2007deep,caldwell2019photonics}.

The description of the excitonic effect deviates significantly from
the single particle response, since to capture the physics of excitons,
many body effects have to be accounted for; this is usually achieved
by solving the Bethe-Salpeter equation (BSE). This integral equation
in momentum space is composed of a kinetic term (obtained from the
single particle response) and an interaction term, which, in general,
couples the electronic degrees of freedom of all the bands of the system via
an electrostatic potential. For the case of an hBN monolayer (or monolayer
TMDs for the same matter), one can simplify the problem by considering
just a single pair of bands, which couple more efficiently than the
remaining ones. This version of the BSE can then be solved using many
methods, with different degrees of numerical complexity, ranging from
fully numerical calculations \citep{Fuchs2013,Komsa2013,Alejandro2016,di2020optical},
to semi-analytical \citep{Pedersen2019,henriques2019optical} and
variational approaches \citep{ditchfield1970self,Zhang2014,quintela2020colloquium}.

A natural extension to the case of a single hBN monolayer, is to consider
the case of bilayers \citep{Ribeiro2011}. 
\begin{figure}[h]
\centering{}\includegraphics[scale=0.9]{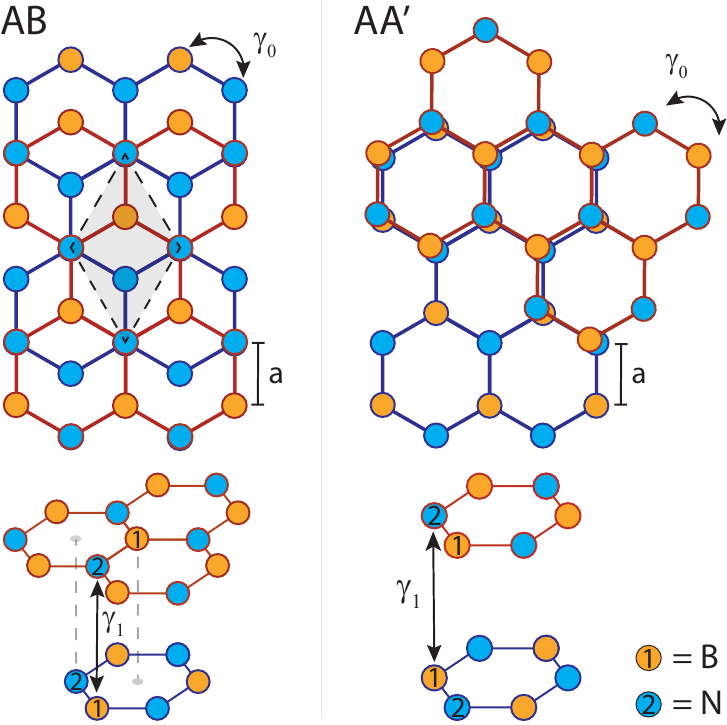}\caption{\label{fig:hBN bilayers}Schematic representation of the lattice of AB and AA' hBN bilayers. The nearest neighbors in-plane and out-of-plane
hoppings are $\gamma_{0}$ and $\gamma_{1}$, respectively. The nearest neighbours distance is $a$.}
\end{figure}
The ground state configuration
for this type of system is the AA' bilayer \citep{Fujimoto2016,wickramaratne2018monolayer},
where the two monolayers are perfectly aligned along the stacking
direction, but the boron and nitrogen atoms sit on opposite sites
in the two planes. Another relevant type of bilayer, with a stability
capable of competing with the AA' configuration, is the AB bilayer,
where two monolayers are shifted relatively to each other. Contrarily
to the monolayers, both of these bilayers present a band gap with
an indirect nature \citep{wickramaratne2018monolayer}, located between
the K point in the valence band and the midway point between the K
and K' points in the conduction band.

Another important aspect regarding the AB and AA' bilayers is that
one configuration can be obtained from the other by a rotation of
$60^\circ$ between the constituent monolayers, making them the limiting cases
of a $0^\circ$ and $60^\circ$ rotation in the study of twisted bilayers. Although
the first theoretical studies on bilayers date back to the time when
graphene was first isolated \citep{JLS2007,JLS2012}, the interest
on the topic only grew since then, remaining an active field of research
at the time of writing \citep{xian2019multiflat,Guinea2021,yao2021enhanced}.
Hence, understanding the optical response of these two configurations
is of great utility to the study of arbitrary twist angles.

Contrarily to the case of hBN monolayers, solving the BSE for the
bilayers is a rather complex process. In fact, this is the reason
why in the current literature this type of problem is almost exclusively treated
with sophisticated numerical approaches \citep{Draxl2018,Ducastelle2018,paleari2018excitons,mengle2019impact,suzuki2020excitons}.
Although accurate, these procedures are rather complex and require
huge computational power. It is clear, then, that a simpler approach
to describe these systems is needed. This is precisely the motivation
behind the current paper, where we study two relevant configurations
for the exploration of twisted hBN bilayers, while presenting a simpler
method to study the excitonic physics in this type of system, with
little computational effort. Even though the excitonic response of the AA' bilayer has already been studied in the literature, the AB bilayer remains largely unexplored.

The text is organized as follows. In Sec. \ref{sec:AB-bilayer} we
consider the case of the AB bilayer, with its study separated into
three stages: first, we study the electronic band structure with a
tight-binding model; then, we introduce the Bethe-Salpeter equation,
and discuss how it can be solved in order to obtain the exciton energies
and wavefunctions; finally, we combine the results of the two previous
stages and evaluate the longitudinal conductivity of the AB configuration
due to excitonic resonances. In Sec. \ref{sec:AA'-bilayer} a similar
analysis is carried out for the AA' bilayer, where the effect of bias
is also discussed. The comparison between the studied configurations
is given in Sec. \ref{sec:Discussion}, where an overview of the work
is also presented. A set of three appendices closes the paper: the first gives the details about density functional theory (DFT) calculations; the second focuses on the discussion of the excitonic angular quantum number; and the last one  describes how to numerical solve the
1D version of the Bethe-Salpeter equation.

\section{AB bilayer\label{sec:AB-bilayer}}

As a starting point to the problem of the excitonic properties of
hBN bilayers, we shall begin by discussing the case of the AB (or
Bernal) configuration. In this type of bilayer one finds that the
two monolayers are shifted relatively to each other along the armchair
direction, as depicted in Fig. \ref{fig:hBN bilayers}.

The first part of our study will be dedicated to the band structure
of such a system, which we will describe using a tight-binding Hamiltonian
and a low energy approximation. Afterwards, the equation that governs
the excitonic problem will be introduced, and a simple approach to
solving it will be presented. At last, the longitudinal conductivity
due to excitons will be evaluated and its main features analyzed,  something which, to the best of our knowledge, remains absent in the literature.

\subsection{Tight binding model}

To characterize the single particle bands of the AB bilayer, let us
start by constructing a minimal tight binding Hamiltonian directly
in momentum space. In our minimal model we account only for nearest
neighbour hoppings, both in the in-plane and out-of-plane directions;
the effect of additional hopping parameters is discussed later in
the text. Following the notation established in Fig. \ref{fig:hBN bilayers},
we consider the basis $\{|1,b\rangle,|2,b\rangle,|2,t\rangle,|1,t\rangle\}$,
where 1 and 2 refer to the sub-lattices (containing boron and nitrogen atoms, respectively), and $b$/$t$ denotes the
bottom/top layer, and find the following Hamiltonian in momentum-space:
\begin{equation}
H_{\textrm{TB},\mathbf{p}}^{AB}=\left[\begin{array}{cccc}
E_{1,b} & \gamma_{0}\phi(\mathbf{p}) & \gamma_{1} & 0\\
\gamma_{0}\phi^{*}(\mathbf{p}) & E_{2,b} & 0 & 0\\
\gamma_{1} & 0 & E_{2,t} & \gamma_{0}\phi^{*}(\mathbf{p})\\
0 & 0 & \gamma_{0}\phi(\mathbf{p}) & E_{1,t}
\end{array}\right],
\end{equation}
where $E_{i,\lambda}$ is the on-site energy of the atom $i$ of the
$\lambda$ layer, $\gamma_{0}$ is the hopping parameter between nearest
neighbors in each monolayer, $\gamma_{1}$ is the interlayer hopping
connecting atoms which are vertically aligned, and $\phi(\mathbf{p})=e^{iap_{y}}+e^{-ia\left(p_{x}\sqrt{3}+p_{y}\right)/2}+e^{ia\left(p_{x}\sqrt{3}-p_{y}\right)/2}$
is a factor which follows from the geometrical configuration
of the lattice, where $a$ is the nearest neighbor distance. Noting that for the AB
configuration we have $E_{1,b}=E_{1,t}$ and $E_{2,b}=E_{2,t}$, we
define $E_{1,b}=E_{g}/2=-E_{2,b}$ to fix the zero of energy.  The $|1,b/t\rangle$ and $|2,b/t\rangle$ sublattices contain boron and nitrogen atoms, respectively. To obtain the values of the different
parameters we fit the energy spectrum of this Hamiltonian to DFT calculations,
the details of which we give in Appendix \ref{sec:Details-on-DFT},
where we also show the tight binding bands fitted to the \emph{ab
initio} results. Doing so we find $E_{g}=4.585$ eV, $\gamma_{0}=2.502$
eV and $\gamma_{1}=0.892$ eV. It is well known that the most common functionals used in DFT underestimate the fundamental band gap, which can be corrected using advanced functionals or $GW$ calculations. This type of approach is, however, beyond
the scope of our work. When the excitonic problem is treated we simply consider the corrected band gap
to be $E_{g}=6.9$ eV, where the band gap correction of \citep{paleari2018excitons} 
was considered (note that even if the correction to the band gap differs
from the one used here, it should not impact the qualitative nature
of the results, and even their quantitative nature should not be drastically
changed). If the limit $\gamma_{1}\rightarrow0$
is considered, we recover a block diagonal Hamiltonian, where each
block describes the electronic properties of a single hBN monolayer,
as expected.

\begin{figure}[h]
\centering{}\includegraphics[scale=0.95]{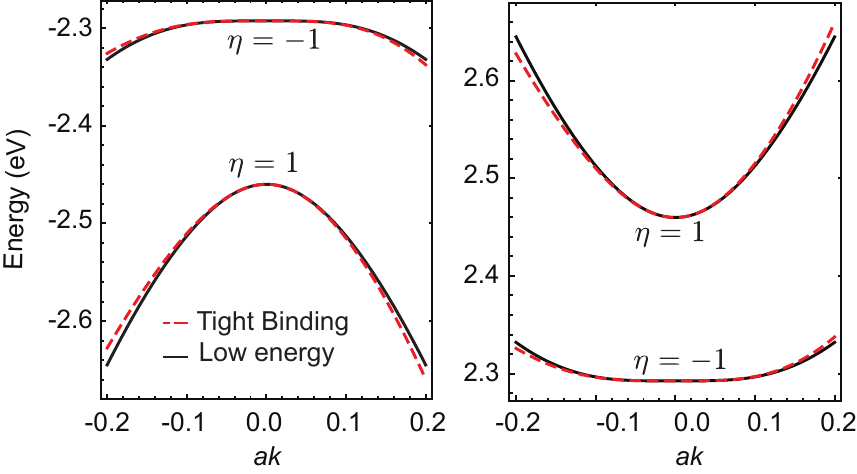}\caption{\label{fig:AB_bands}Valence and conduction bands obtained from the
tight binding Hamiltonian, and the low energy approximation for the
AB bilayer. The momenta is measured relatively to the K point (the
results near the K' point are identical). The results obtained from
the two approaches are in total agreement.}
\end{figure}

Since we will be mostly interested on the low energy optical response,
we restrict our analysis to the Dirac valleys, that is, the region
around the vertices of the first Brillouin zone (1BZ), also known
as the $K/K'$ points. To do this, we write $\mathbf{p} =\tau \mathbf{K} + \mathbf{k}$ and approximate $\phi(\tau \mathbf{K} + \mathbf{k})$ to first order in $\mathbf{k}$
as $\phi_{\tau}(\mathbf{k})\approx-\frac{3}{2}a\left(\tau k_{x}-ik_{y}\right)$,
with $\tau=\pm1$ labeling the $K/K'$ points respectively. Notice
that, hereinafter, the values of $\mathbf{k}=(k_{x},k_{y})$ are measured
relatively to these points in the reciprocal space. With this approximation
one finds the following low energy Hamiltonian
\begin{align}
H_{\textrm{low},\mathbf{k}}^{AB} & =\sigma_{+}\otimes\left[\hbar v_{F}\left(\tau k_{x}\sigma_{x}+k_{y}\sigma_{y}\right)+\frac{E_{g}}{2}\sigma_{z}\right]\nonumber \\
 & +\sigma_{-}\otimes\left[\hbar v_{F}\left(\tau k_{x}\sigma_{x}-k_{y}\sigma_{y}\right)-\frac{E_{g}}{2}\sigma_{z}\right]\nonumber \\
 & +\sigma_{x}\otimes\sigma_{+}\gamma_{1},
\end{align}
where $\sigma_{\pm}=\left(I\pm\sigma_{z}\right)/2$ and $\hbar v_{F}=3\gamma_{0}a/2$.
Diagonalizing this Hamiltonian we find the energy dispersion relations
\begin{align}
E_{\mathbf{k}}^{\lambda,\eta} & =\frac{\lambda}{2}\sqrt{E_{g}^{2}+4\hbar^{2}v_{F}^{2}k^{2}+2\gamma_{1}^{2}+2\gamma_{1}\eta\Lambda_{\mathbf{k}}}
\end{align}
with $\Lambda_{\mathbf{k}}=\sqrt{\gamma_{1}^{2}+4\hbar^{2}v_{F}^{2}k^{2}}$,
$\lambda=\pm1$ (when used as a number) or $\lambda=c/v$ (when used
as an index) and $\eta=\pm1$. Just like in the case of an hBN monolayer,
the energy spectrum is the same for $\tau=1$ or $\tau=-1$, since
we ignore the small effect of spin-orbit coupling in this system.
In Fig. \ref{fig:AB_bands} we depict the band structure obtained
from the tight binding model as well as from the low energy approximation
in the vicinity of the Dirac points; the agreement between the two
results is clear, as it should. 
Moreover we find that the bands associated with the index $\eta=+1$
take an approximately parabolic shape, while those with $\eta=-1$
present a momentum dependence proportional to $k^{4}$. A similar
band structure is found on bilayer TMDs, such as 3R-MoS\textsubscript{2}
\citep{paradisanos2020controlling}. The eigenvectors associated with
each band read
\begin{align}
|u_{\mathbf{k}}^{v,\tau,\eta}\rangle & =\frac{1}{\sqrt{\mathcal{V}_{\eta,\mathbf{k}}}}\left[\begin{array}{c}
\tau e^{\tau i\theta}\frac{\gamma_{1}+\eta\Lambda_{\mathbf{k}}}{2\hbar v_{F}k}\\
e^{2\tau i\theta}\frac{\left(E_{g}-2E_{\mathbf{k}}^{v,\eta}\right)\left(\gamma_{1}-\eta\Lambda_{\mathbf{k}}\right)}{4k^{2}\hbar^{2}v_{F}^{2}}\\
-\tau e^{\tau i\theta}\frac{E_{g}-2E_{\mathbf{k}}^{v,\eta}}{2\hbar v_{F}k}\\
1
\end{array}\right],\\
|u_{\mathbf{k}}^{c,\tau,\eta}\rangle & =\frac{1}{\sqrt{\mathcal{C}_{\eta,\mathbf{k}}}}\left[\begin{array}{c}
\tau e^{\tau i\theta}\frac{\gamma_{1}+\eta\Lambda_{\mathbf{k}}}{2\hbar v_{F}k}\\
e^{2\tau i\theta}\frac{\left(E_{g}-2E_{\mathbf{k}}^{c,\eta}\right)\left(\gamma_{1}-\eta\Lambda_{\mathbf{k}}\right)}{4k^{2}\hbar^{2}v_{F}^{2}}\\
-\tau e^{\tau i\theta}\frac{E_{g}-2E_{\mathbf{k}}^{c,\eta}}{2\hbar v_{F}k}\\
1
\end{array}\right].\label{eq:AB_low_energy_spinors}
\end{align}
where $\mathcal{C}_{\eta,\mathbf{k}}/\mathcal{V}_{\eta,\mathbf{k}}$
are normalization factors and $\theta=\arctan(k_{y}/k_{x})$. Just
like for any other state vector, these spinors are defined up to a
global phase factor (for example $e^{\tau i\theta}$). 
The particular choice used in Eq. II.5 was made in order to simplify the numerical formulation of the excitonic problem, which will be discussed in the following section.
At last, let us note for future reference that for small
momentum these vectors take the approximate form
\begin{align}
|u_{\mathbf{k}}^{v,-}\rangle & \approx\left[\begin{array}{c}
0\\
e^{2\tau i\theta}\\
0\\
0
\end{array}\right],\qquad|u_{\mathbf{k}}^{v,+}\rangle\approx\left[\begin{array}{c}
e^{i\tau\theta}\sin\frac{\xi}{2}\\
0\\
-e^{i\tau\theta}\cos\frac{\xi}{2}\\
0
\end{array}\right],\\
|u_{\mathbf{k}}^{c,-}\rangle & ,\approx\left[\begin{array}{c}
0\\
0\\
0\\
1
\end{array}\right]\qquad|u_{\mathbf{k}}^{c,+}\rangle\approx\left[\begin{array}{c}
e^{i\tau\theta}\cos\frac{\xi}{2}\\
0\\
e^{i\tau\theta}\sin\frac{\xi}{2}\\
0
\end{array}\right]\label{eq:small momentum u},
\end{align}
with $\xi=\arctan\left[2\gamma_{1}/E_{g}\right]$.

\subsection{Excitonic problem}

Now that the single particle bands and Bloch factors were determined, let us tackle the problem of obtaining the excitonic energies and wavefunctions.

To obtain the energies and wave functions of the excitons in the AB
bilayer we shall solve the well known Bethe-Salpeter equation (BSE).
The BSE is an integral equation in momentum space, which requires
the information of the single particle approximation, and whose solution
determines the excitonic spectrum. Explicitly, for an exciton with zero center of mass momentum, this equation reads:
\begin{align}
\left(E_{\mathbf{k}}^{c}-E_{\mathbf{k}}^{v}\right)\psi_{cv}(\mathbf{k})-\sum_{\mathbf{q}c'v'}V(\mathbf{k}-\mathbf{q})\langle u_{\mathbf{k}}^{c}|u_{\mathbf{q}}^{c'}\rangle\langle u_{\mathbf{q}}^{v'}|u_{\mathbf{k}}^{v}\rangle\nonumber \\
\times\psi_{c'v'}(\mathbf{q})=E\psi_{cv}(\mathbf{k}),\label{eq:BSE}
\end{align}
where, for the sake of a simpler notation, we have omitted the indexes
$\tau$ and $\eta$, which are now included in the band index ($c/c'$
or $v/v'$). Here, the sum is performed over the momentum $\mathbf{q}$
and all the bands of our model; $\psi_{cv}(\mathbf{k})$ refers to
the exciton's wave function projected onto the pair of bands $(v,c)$,
$E$ corresponds to the exciton's energy and $V(\mathbf{k}-\mathbf{q})$ is the Fourier transform of the electron-hole interaction, which we model with the Rytova-Keldysh potential \citep{rytova1967,keldysh1979coulomb,Cudazzo2011}.
This potential can be obtained from the solution of the Poisson equation
for a charge embedded in a thin film, and is known to accurately capture
the electrostatic interaction in 2D materials; it reads
\begin{equation}
V(\mathbf{k})=\frac{\hbar c\alpha}{\epsilon k\left(1+r_{0}k\right)},\label{eq:V_RK (k)}
\end{equation}
with $c$ the speed of light, $\alpha\sim1/137$ the fine structure
constant, $\epsilon$ the mean dielectric constant of the media above
and below the monolayer and $r_{0}$ an in-plane screening length,
which is related with the 2D polarizability of the system \citep{tian2019electronic}.
We now note that Eq. (\ref{eq:BSE}) corresponds, in fact, to a set
of four coupled equations, one for each pair of valence and conduction bands
$(v,c)$, defining an eigenvalue problem. In this type of system,
the formation of an exciton can not be \emph{a priori} assigned to a single pair of
bands, but rather to a cooperative process where the four bands of
the model contribute to the formation of such an entity. 
Furthermore, from
Eq. (\ref{eq:BSE}), one already sees that the phases chosen for the
Bloch factors in Eq. (\ref{eq:AB_low_energy_spinors}) have an impact on the BSE, since different phase choices
lead to different angular dependencies for the term $\langle u_{\mathbf{k}}^{c}|u_{\mathbf{q}}^{c'}\rangle\langle u_{\mathbf{q}}^{v'}|u_{\mathbf{k}}^{v}\rangle$.
We stress, however, that when a physical quantity is computed, for
example a conductivity, its final result is independent from the phase
one initially chose for the Bloch factors.

Solving the BSE is no simple task, and, as mentioned in the introduction,
different techniques are frequently employed to achieve this. The approach we consider here is to use the results of the tight binding
model we previously presented, and to reduce the BSE to a 1D integral
equation, which can then be easily solved with a single numerical
quadrature. In what follows we give a brief description of the approach
we use, with a more detailed technical discussion presented in Appendix
\ref{sec:Solving-the-BSE}.

The first step to transform the BSE into a 1D integral equation is
to consider the system to be isotropic, which allows us to
write the exciton's wave function as the product of a radial and an
angular components, such as $\psi_{cv}(\mathbf{k})=f_{cv}(k)e^{im\theta_{k}}$,
with $m$ an integer. At first, one might be tempted to associate
the value of $m$ with the angular momentum of the exciton, however
this is not necessarily true. From the study of hBN monolayers (or
other systems which can be treated with a two band model), it is known
that the number which characterizes the angular momentum is obtained
from a combination of the $m$ present in the envelop function $\psi_{cv}(\mathbf{k})$
with an additional contribution stemming from the pseudospin of the
system \citep{park2010tunable,Cao2018}. However, for a model with
four bands (like the one we currently consider) the identification
of the pseudospin contribution is unclear, and because of that we
will refrain from attributing an angular quantum number to excitons
that appear from the solution of the BSE when the four bands are accounted
for. In Appendix \ref{sec:On-the-exciton's} we give a more detailed
discussion on this.

Making use of the above mentioned proposal for the wave function $\psi_{cv}(\mathbf{k})=f_{cv}(k)e^{im\theta_{k}}$,
the BSE acquires the form:
\begin{align}
\left(E_{k}^{c}-E_{k}^{v}\right)f_{cv}(k)-\sum_{c'v'}\int qdqd\theta_{q}V(\mathbf{k}-\mathbf{q})\langle u_{\mathbf{k}}^{c}|u_{\mathbf{q}}^{c'}\rangle\langle u_{\mathbf{q}}^{v'}|u_{\mathbf{k}}^{v}\rangle\nonumber \\
\times f_{c'v'}(q)e^{im\left(\theta_{q}-\theta_{k}\right)}=Ef_{cv}(k).\label{eq:simpler_BSE}
\end{align}
We now note that according to Eq. (\ref{eq:V_RK (k)}), $V(\mathbf{k}-\mathbf{q})$
is a function of $k$, $q$ and $\cos(\theta_{q}-\theta_{k})$, that
is $V(\mathbf{k}-\mathbf{q})\equiv V(k,q,\theta_{q}-\theta_{k})$.
Knowing this, one easily sees that if the angular dependence of the
spinor product $\langle u_{\mathbf{k}}^{c}|u_{\mathbf{q}}^{c'}\rangle\langle u_{\mathbf{q}}^{v'}|u_{\mathbf{k}}^{v}\rangle$
only contains terms of the form $e^{in(\theta_{q}-\theta_{k})},$
with $n$ a real number, then the integral over $d\theta_{q}$ can
be converted into an integral over a new variable $\vartheta=\theta_{q}-\theta_{k}$,
independent of $q$ and $k$. By removing the momentum dependence
from the angular integral, its evaluation can be thought of as an
independent step of the calculation, thus effectively transforming
the BSE into a 1D integral equation (whose only integration variable
is now $q$), which can then be easily solved (see Appendix \ref{sec:Solving-the-BSE}). This approach 
is computationally advantageous when compared with a strictly two
dimensional calculation (which scales as $N^{4}$ while the simpler
1D problem scales as $N^{2}$, with $N$ the number of points in the numerical quadrature).

The key point now is to find the spinor's phase choice which guarantees
that their product has the desired angular dependence. First, we note
that for the term $\langle u_{\mathbf{k}}^{c}|u_{\mathbf{q}}^{c'}\rangle\langle u_{\mathbf{q}}^{v'}|u_{\mathbf{k}}^{v}\rangle$
with $c=c'$ and $v=v'$, the angular dependence always presents the
form we are seeking, regardless of the phase choice, since the phase
of each $|\textrm{ket}\rangle$ is balanced by the phase of the $\langle\textrm{bra}|$
with which it is contracted. This is precisely what one finds in the
case of monolayers, where the BSE can consistently be transformed
into a 1D integral equation \citep{Pedersen2019}. What about the
remaining terms where $c\neq c'$ and/or $v\neq v'$? Depending on
the phase choice for the spinors one may find that unwanted terms,
such as $e^{in\theta_{q}}e^{-ip\theta_{k}},$with $p\neq n$, appear.
Using Eq. (\ref{eq:AB_low_energy_spinors}), however, produces the
desired angular dependence for all the products of spinors that appear
in the BSE, thus allowing us to convert the excitonic problem into
a 1D integral equation. 

Using the method we have just now highlighted, and discuss in more
detail in Appendix \ref{sec:Solving-the-BSE}, we solved the BSE for
the AB bilayer for different values of $m$ (which, we recall, does not correspond directly to the angular quantum number). We considered the bilayer to be suspended,
$\epsilon=1$, and used $r_{0}=16\text{Å}$ in agreement with the
value found from \emph{ab initio} calculations in \citep{paleari2018excitons}. When solving the BSE we employed a Gauss-Legendre quadrature,
containing 100 points, which we verified to be more than enough to guarantee
the convergence of the energies and wave functions for the first ten excitonic states.

\begin{figure}[h]
\centering{}\includegraphics[scale=0.82]{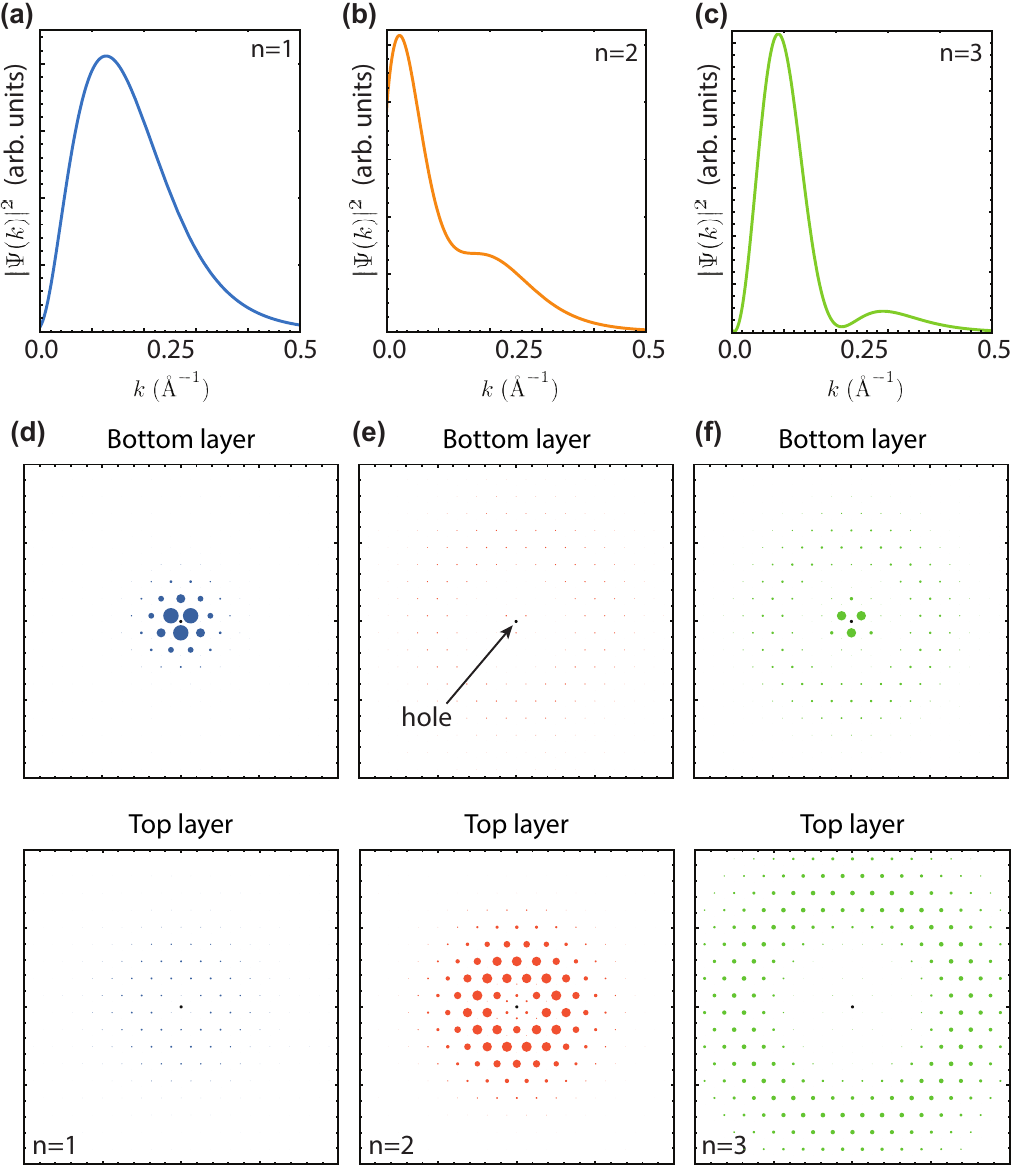}\caption{\label{fig:AB_WF_Panel}(a)-(c): Wave functions in momentum space,
$|\boldsymbol{\Psi}(k)|^{2}=\sum_{cv}|\psi_{cv}(k)|^{2}$, of the
three excitonic states which are responsible for the first three resonances
of the optical conductivity. Their energies are $E_{n=1}=5.71$ eV,
$E_{n=2}=6.13$ eV and $E_{n=3}=6.34$ eV. (d)-(f) Real space representation
of the absolute value of the wave function in both layers when the hole is placed
on the nitrogen site of the bottom layer ($|2,b\rangle$ sub-lattice).}
\end{figure}

In panels (a) to (c) of Fig. \ref{fig:AB_WF_Panel}, we depict the
wave functions, $|\Psi(k)|^{2}=\sum_{cv}|\psi_{cv}(k)|^{2}$, associated
with three of the states found from the solution of the BSE. 
These states are some of the most relevant ones for the linear optical
response of the system (computed in the following section), and their
energies read $E_{n=1}=5.71$ eV, $E_{n=2}=6.13$ eV and $E_{n=3}=6.34$
eV. Analyzing the three panels, we see that the $n=1$
and $n=3$ states present wave functions which are similar to those
found in the bound states of the 2D Hydrogen atom \citep{Chao1991}
(which in turn are similar to those of its three dimensional counterpart).
In fact, since these wave functions are zero at the origin, have an
approximately linear behavior for small momentum, and present zero
and one nodes, respectively, they bare a particular resemblance with
the 2p and 3p states of the Hydrogen atom. At odds with this, the
wave function of the $n=2$ state presents a more exotic behavior,
with a broad shoulder instead of a node, unlike an Hydrogenic wave
function.

To gain more information about these states, especially regarding
their configuration in real space, we compute the projection of their
wave functions onto the electron and hole sub-lattices, which can
be written as:
\begin{equation}
\Psi_{\alpha\beta}(\mathbf{r}_{e},\mathbf{r}_{h})=\sum_{\mathbf{k},c,v}e^{i(\boldsymbol{K}+\mathbf{k})\cdot\left(\mathbf{r}_{e}-\mathbf{r}_{h}\right)}\psi_{cv}(\mathbf{k})u_{\mathbf{k},c}^{\alpha}\left(u_{\mathbf{k},v}^{\beta}\right)^{*},
\end{equation}
where $\mathbf{r}_{e}$ and $\mathbf{r}_{h}$ are the electron and
hole positions, respectively, and $u_{\mathbf{k},c}^{\alpha}$ refers
to the $\alpha$ sub-lattice entry of the Bloch factor $|u_{\mathbf{k},c}\rangle$
(an analogous definition holds for $u_{\mathbf{k},v}^{\beta}$). For
simplicity we consider $\mathbf{r}_{h}=0$ and study the behavior
of the wave function with $\mathbf{r}_{e}$. Notice how the the term $\boldsymbol{K}+\mathbf{k}$ appears on the complex exponential because the momenta are being measured relatively to the Dirac point; however, the contribution from $\boldsymbol{K}$ vanishes when the square modulus of the wave function is considered. In Fig. \ref{fig:AB_WF_Panel} (d)-(f) we depict the real space wave functions when the hole is place on the nitrogen atom of the bottom layer ($|2,b\rangle$); the position of the hole is marked by a small black dot in the center of each figure. For the $n=1$ exciton, we find that the wave function is mainly distributed on the bottom layer boron sites (this is the reason we apparently only see a triangular lattice, instead of a honeycomb one), that is, on the same layer as the hole, with a smaller portion being present on the top layer; this distribution of the wave function indicates that this state has a predominantly intralayer nature. On the other hand, for the $n=2$ and $n=3$ excitons, we find a rather significant part of the wave function spread over the top layer, indicating the interlayer character of these excitations. To more easily understand how the wave function behaves for different positions of the hole, we present in Table \ref{tab:AB_real_space_WF} the values found for the integrated square modulus of the wave function, $\int | \Psi_{\alpha \beta} (\mathbf{r}_e,0)|^2 d\mathbf{r}_e$, which gives the probability of finding the electron on one of the layers, for each possible location of the hole.
\begin{table}[h]
\begin{tabular}{|l|l|l|l|l|l|}
\hline
                                             &        & $|1,b\rangle$ & $|2,b\rangle$ & $|2,t\rangle$ & $|1,t\rangle$ \\ \hline
\multicolumn{1}{|c|}{\multirow{2}{*}{$n=1$}} & Bottom & 0.01          & 0.34          & 0.03          & 0.00          \\ \cline{2-6} 
\multicolumn{1}{|c|}{}                       & Top    & 0,00          & 0.26          & 0.34          & 0.01          \\ \hline
\multirow{2}{*}{$n=2$}                       & Bottom & 0.00          & 0.17          & 0.01          & 0.00          \\ \cline{2-6} 
                                             & Top    & 0.02          & 0.61          & 0.17          & 0.01          \\ \hline
\multirow{2}{*}{$n=3$}                       & Bottom & 0.01          & 0.17          & 0.08          & 0.00          \\ \cline{2-6} 
                                             & Top    & 0.00          & 0.57          & 0.16          & 0.01          \\ \hline
\end{tabular}
\caption{\label{tab:AB_real_space_WF} Integrate values of the square modulus of the sublattice resolved real space wave function for the AB bilayer. The horizontal top row indicates the sublattice where the hole is placed, while the two columns on the left indicate the exciton we are considering, and the layer where the electron is located.}
\end{table}
From the inspection of this table, one finds that: i) there is a clear preference for the hole to be located on the $|2,t/b\rangle$ sublattices (containing nitrogen atoms), given the small values found for the integrated wave function when the hole is located on either $|1,b/t\rangle$ sublattices; ii) we confirm the previous assignment of the $n=1$ exciton as mainly intralayer, while the $n=2$ and $n=3$ ones are mostly interlayer.

\subsection{Optical conductivity}

\begin{figure}[h]
\centering{}\includegraphics[scale=1.05]{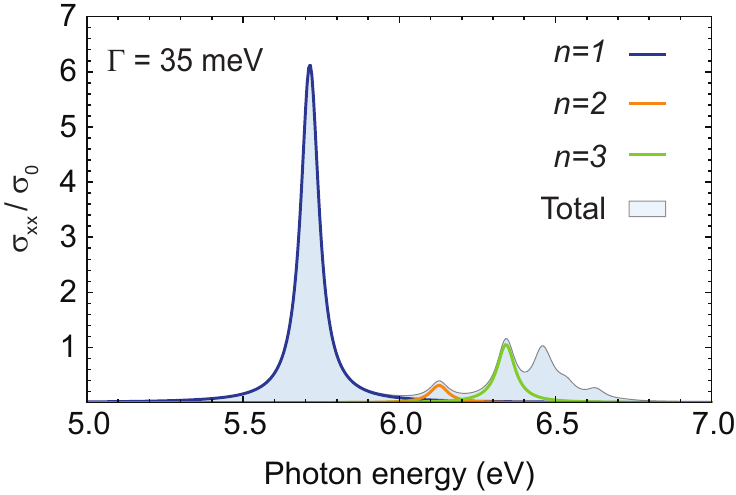}\caption{\label{fig:AB_sigma_xx}Optical conductivity of an AB hBN bilayer accounting for the first 10 exciton states with $m=\tau$ (the contributions from both valleys were accounted for).
The first three resonances correspond to the excitonic states depicted
in Fig. \ref{fig:AB_WF_Panel}. To obtain this result a phenomenological
broadening of 35 meV was considered. The conductivity is given in
terms of the conductivity of graphene $\sigma_{0}=e^{2}/4\hbar$.}
\end{figure}

Now that the the BSE was solved for the AB bilayer, we are ready to
evaluate its conductivity due to the excitonic effect. Following Ref.
\citep{Thomas2015cond}, we write the conductivity for a multiband
system as
\begin{equation}
\frac{\sigma(\omega)}{\sigma_{0}}=\frac{i}{\pi}\sum_{n}E_{n}\frac{\boldsymbol{\Omega}_{n}\boldsymbol{\Omega}_{n}^{*}}{E_{n}-\hbar\omega}+(\omega\rightarrow-\omega)^{*},\label{eq:Thomas_cond}
\end{equation}
where $\sigma_{0}=e^{2}/4\hbar$ is the conductivity of graphene,
the sum over $n$ runs over the different exciton states with energy
$E_{n}$ and
\begin{equation}
\boldsymbol{\Omega}_{n}=\sum_{vc\mathbf{k}}\psi_{cv}^{(n)}(\mathbf{k})\boldsymbol{\Omega}_{vc\mathbf{k}}
\end{equation}
where $\boldsymbol{\Omega}_{vc\mathbf{k}}$ is the position operator
interband matrix element, which we write as
\begin{equation}
\boldsymbol{\Omega}_{vc\mathbf{k}}=\frac{\langle u_{\mathbf{k}}^{v}|\left[H,\mathbf{r}\right]|u_{\mathbf{k}}^{c}\rangle}{E_{\mathbf{k}}^{v}-E_{\mathbf{k}}^{c}},
\end{equation}
with $H$ standing for the low energy tight binding Hamiltonian. The
evaluation of the interband matrix element is crucial to determine
which of the solutions of the BSE couple with the electric field,
and consequently contribute to the conductivity. For the current system,
the interband matrix elements imposes that only states with $m=\pm \tau$
may give a finite contribution (we recall once more that this does not correspond to the angular quantum number). Not only that, but the sum over the bands
also plays a role in determining which states couple more efficiently
with light due to the possibility of existing constructive or destructive
interference between the different terms. Using the solutions of the
BSE given in the previous section, we compute the optical response
of the system due to a linearly polarized electric field; its conductivity
is depicted in Fig. \ref{fig:AB_sigma_xx}, where a phenomenological
broadening of 35 meV was considered for all resonances.
The shaded blue area corresponds to the conductivity accounting for
10 exciton states (all with $m=\tau$, since we found the $m=-\tau$ states to have rather small oscillator strengths); the contributions of the states highlighted in
Fig. \ref{fig:AB_WF_Panel} are depicted in the same color as the
corresponding wave functions.. From this figure, we see that the longitudinal
conductivity of the AB hBN bilayer has its more pronounced feature
on the first resonance, while a set of lower intensity ones appear
at higher energies. Furthermore, the conductivity of the AB bilayer
presents a small, yet noticeable, resonance between the first and
third peaks, which can be ascribed to the second state of Fig. \ref{fig:AB_WF_Panel}.
 Above the third resonance and up to approximately 6.5 eV, three resonances appear. These peaks, however, overlap significantly, making it difficult to resolve them. Moreover, since our model is based on a low energy approximation, the results are expected to become progressively less accurate as we approach the band edge. Due to these two reasons we focus our analysis solely on the first three resonances.
%

\subsection{Exciton angular quantum number \label{subsec:Complementary-analysis}}

Having determined the complete longitudinal conductivity using the
results of the four band BSE, we shall now carry out a complementary
analysis to gain further insight on the nature of each resonance,
especially regarding the angular quantum number of the excitons behind
them.

As a first, and somewhat naive, approach, we return to the BSE and
restrict it to a single pair of valence and conduction bands. In particular,
we consider only the bands which present an energy dispersion in $k^{4}$,
since intuition tells us that these should dominate in the low energy
response. Because in this approximation we are effectively treating
a two band problem, we can identify the contribution of the pseudo-spin
to the angular quantum number (see Appendix \ref{sec:On-the-exciton's}).

Let us define the excitonic wave function in real space for a two band model \citep{park2010tunable} as
\begin{align}
	\Psi_{\alpha,\beta} (\mathbf{r}_e,\mathbf{r}_h) = \sum_{\mathbf{k}} e^{i (\boldsymbol{K}+\mathbf{k}) \cdot (\mathbf{r}_e - \mathbf{r}_h)} \psi_{cv} (\mathbf{k})  u^\alpha_{\mathbf{k},c} \left( u^\beta_{\mathbf{k},v} \right)^*
\end{align}
which is analogous to the previously given definition, only this time without the sum over the bands, since a single pair is being considered. From Eq. (\ref{eq:small momentum u}), one sees that for small momentum the product $u^\alpha_{\mathbf{k},c} \left( u^\beta_{\mathbf{k},v} \right)^*$ approximately introduces an additional phase of $e^{-2i\tau\theta}$ (recall that only the bands with $\eta=-1$ are being currently  considered), which can be combined with the angular part of $\psi_{cv}(\mathbf{k})$. Hence, within this approximation,
we may define the angular quantum number of the exciton as $m_{\textrm{X}}=m+m_{\textrm{ps}}$
where $m_{\textrm{ps}}=-2\tau$ is the pseudo-spin contribution to
the angular quantum number, and $m$ is the contribution from the envelope function $\psi_{cv}(\mathbf{k})$.

When the conductivity is evaluated, the interband matrix element imposes
that only states with $m=\pm\tau$ may couple with the external excitation.
Thus, taking into consideration the definition of the angular quantum
number $m_{\textrm{X}}$, we find that, at least approximately, only
states with angular quantum numbers $m_{\textrm{X}}=-\tau$ or $m_{\textrm{X}}=-3\tau$
are optically bright. In analogy with the Hydrogen atom, we label
these states as $p$- and $f$-states, since the modulus of their
angular quantum number is 1 and 3, respectively. These selection rules
are in line with the momentum space wave functions depicted in Fig.
\ref{fig:AB_WF_Panel}.

\begin{figure}[h]
\centering{}\includegraphics[scale=0.95]{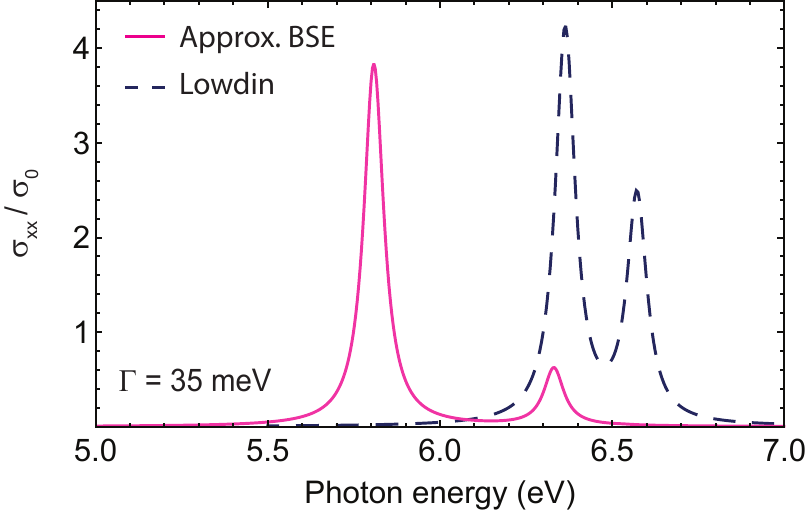}\caption{\label{fig:AB_approx_sigma_xx}Optical conductivity of an AB hBN bilayer
obtained by: i) considering only two of the original four bands when
solving the BSE (labeled as Approx. BSE); ii) using the Lowdin partitioning.
As in Fig. \ref{fig:AB_sigma_xx} a broadening of 35 meV was introduced.
The conductivities are given in terms of the conductivity of graphene
$\sigma_{0}=e^{2}/4\hbar$.}
\end{figure}

In Fig. \ref{fig:AB_approx_sigma_xx}, we depict the conductivity
found with this two band approximation, where once again a phenomenological
broadening of 35 meV was considered; only the first two $p-$states
were accounted for since the $f-$states appear above these two, and
with a far smaller oscillator strength.
Comparing this result with the one of Fig. \ref{fig:AB_sigma_xx},
one clearly sees the resemblance between the two conductivities, both
in the location of the resonances as well as their relative magnitude.
The absolute magnitude, is slightly different from what was found
when the four band BSE was solved; this is to be expected, since in
the current approximation we are neglecting the contribution of other
pairs of bands to the conductivity. Thus, it appears that one can
confidently assign, at least approximately, the Hydrogenic labels of 2p and 3p
states to the excitons which originate the first and third resonances
of the conductivity in Fig. \ref{fig:AB_sigma_xx}. Note, however,
how the small resonance at approximately 6 eV in Fig. \ref{fig:AB_sigma_xx} is absent
in this approximation. By repeating this procedure for all possible
pairs of bands, we find that using the $\eta=-1$ bands gives the best results when compared with the four band calculation.  Moreover, we note that the small resonance is only ever captured when the four bands are accounted for, indicating a clear difference of this exciton when compared with the other two we are considering (which can be approximately captured by selecting two of the four bands of our model).

To further confirm the correct labeling of the resonances we can follow
the ideas of Ref. \citep{Zhang2018}, where the process of folding
a tight binding Hamiltonian on itself, i.e. applying a Lowdin partitioning
\citep{lowdin1951note,winkler2003spin}, was used to obtain the optical
selection rules of a 3R-MoS\textsubscript{2} bilayer. In a succinct
manner, to obtain an effective $2\times2$ Hamiltonian from a given
$4\times4$ model Hamiltonian, one should start by finding the unitary
transformation which diagonalizes the model Hamiltonian at $k=0$.
Then, the unitary transformation should be applied to the model Hamiltonian
with finite $k$, and the basis should be reordered such that the
low energy diagonal terms appear on the upper left $2\times2$ block.
At last, the effective Hamiltonian is obtained from this one through
the relation
\begin{equation}
\left(H_{\textrm{eff}}\right)_{ij}=\tilde{H}_{ij}+\frac{1}{2}\sum_{l}\frac{\tilde{H}_{il}\tilde{H}_{lj}}{H_{ii}-H_{ll}}+\frac{\tilde{H}_{il}\tilde{H}_{lj}}{H_{jj}-H_{ll}},\label{eq:Lowdin}
\end{equation}
with $i,j=\{1,2\}$ and $l=\{3,4\}$; $\tilde{H}$ corresponds to
the model Hamiltonian after applying the unitary transformation and
rearranging its basis. Hence, using the described procedure to project
the high energy bands onto the low energy ones, we obtain the following
effective two band Hamiltonian:
\begin{equation}
H_{\textrm{eff}}\approx\left(\begin{array}{cc}
\frac{E_{g}}{2} & -\frac{\hbar^{2}v_{F}^{2}}{\gamma_{1}}k^{2}e^{-2i\tau\theta}\\
-\frac{\hbar^{2}v_{F}^{2}}{\gamma_{1}}k^{2}e^{2i\tau\theta} & -\frac{E_{g}}{2}
\end{array}\right).
\end{equation}
According to Ref. \citep{Zhang2018}, the winding number associated
with this Hamiltonian is $w=-2\tau$; and the optical selection rules
follow from the winding number as $m_X=w\pm\tau$, when trigonal warping
is neglected. Thus, using this alternative approach, we once again
find selection rules which only allow the excitation of states with
$m_X=-\tau$ and $m_X=-3\tau$, that is, p and f-states. If the effect
of trigonal warping had been included, for example by introducing
hopping to second neighbors (either in the in-plane or out of plane
directions), the set of selection rules would be extended to include
s- and d-states (with angular quantum number 0 and 2, respectively), due to an additional contribution of a factor of 3 to $m_X$ stemming from the symmetry of the lattice.
Since the resonances associated with these states would be
proportional to the square of the associated hopping integral, which
is significantly smaller than the nearest neighbors hoppings, their
intensity would be rather small when compared to the resonances we
have accounted for here.

At last, we note that if one now solves the excitonic problem using
$H_{\textrm{eff}}$ as a starting point, the exciton energies will
be significantly overestimated. This is a consequence of the band
structure given by $H_{\textrm{eff}}$, where the dispersion relation
presents a $k^{4}$ dependence near $k\rightarrow0$, but grows at
much faster rate than the original bands as the momentum increases.
This results in a higher kinetic energy for the electrons, which in
turn reduces the exciton binding energies. Hence, even though the
Lowdin partitioning captures the qualitative features of the conductivity
of the AB bilayer, and easily gives optical selection rules, it
fails to quantitatively describe its conductivity, as we show in Fig.
\ref{fig:AB_approx_sigma_xx}.

\section{AA' bilayer\label{sec:AA'-bilayer}}

In this section we will focus on AA' bilayers. In this type of bilayer, the two monolayers
are vertically aligned, with the boron and nitrogen atoms in opposite
sites, such that a boron/nitrogen atom is always vertically aligned
with a nitrogen/boron atom. A depiction of this type of bilayer is
presented in Fig. \ref{fig:hBN bilayers}. 
As in the previous section, we will begin by studying the electronic
band structure of the system followed by the calculation of the excitonic
response. Since the ideas and techniques of the previous section carry
on to the current one, in what follows we will give a less detailed
description on how the results were obtained, and will mainly focus
on the differences between the two types of bilayer.

\subsection{Tight binding model}

To obtain the low energy band structure of the AA' bilayer we will
once more use a tight binding Hamiltonian written directly in momentum
space. Working in the basis $\{|1,b\rangle,|2,b\rangle,|1,t\rangle,|2,t\rangle\}$
(see Fig. \ref{fig:hBN bilayers}), we write
\begin{equation}
H_{\textrm{TB},\mathbf{p}}^{AA'}=\left[\begin{array}{cccc}
E_{g}/2 & \gamma_{0}\phi^{*}(\mathbf{p}) & 0 & \gamma_{1}\\
\gamma_{0}\phi(\mathbf{p}) & -E_{g}/2 & \gamma_{1} & 0\\
0 & \gamma_{1} & E_{g}/2 & \gamma_{0}\phi(\mathbf{p})\\
\gamma_{1} & 0 & \gamma_{0}\phi^{*}(\mathbf{p}) & -E_{g}/2
\end{array}\right].
\end{equation}
Here we have considered $E_{1,b}=E_{1,t}$ and $E_{2,b}=E_{2,t}$,
and defined $E_{g}$ as $E_{1,b}=E_{g}/2=-E_{2,b}$. The $|1,b/t\rangle$ and $|2,b/t\rangle$ contain boron and nitrogen atoms, respectively. As in the AB
bilayer, $\gamma_{0}$ and $\gamma_{1}$ refer to the intra and interlayer
nearest neighbors hoppings, respectively, and $\phi(\mathbf{p})$
is a phase factor whose expression is the same as in the previous
section. Notice how for the AA' bilayer the Hamiltonian presents twice
as many $\gamma_{1}$ than for the AB configuration, in agreement
with the increased number of atoms which are vertically aligned. As
before, if $\gamma_{1}\rightarrow0$, one is left with a block diagonal
Hamiltonian describing two decoupled monolayers. To obtain the numerical
values for the different parameters of the model, the energy spectrum
of the tight binding Hamiltonian was fitted to DFT calculations (obtained
in an identical manner to what was described in the previous section),
yielding $E_{g}=4.65$ eV, $\gamma_{0}=2.491$ eV and $\gamma_{1}=0.595$
eV.

In the low energy approximation, that is, near the Dirac points, we
write $\phi_{\tau}(\mathbf{k})\approx-\frac{3}{2}a\left(\tau k_{x}-ik_{y}\right)$,
and find the effective low energy Hamiltonian
\begin{align}
H_{\textrm{low},\mathbf{k}}^{AA'} & =\sigma_{+}\otimes\left[\hbar v_{F}\left(\tau k_{x}\sigma_{x}-k_{y}\sigma_{y}\right)+\frac{E_{g}}{2}\sigma_{z}\right]\nonumber \\
 & +\sigma_{-}\otimes\left[\hbar v_{F}\left(\tau k_{x}\sigma_{x}+k_{y}\sigma_{y}\right)+\frac{E_{g}}{2}\sigma_{z}\right]\nonumber \\
 & +\sigma_{x}\otimes\sigma_{x}\gamma_{1}\label{eq:AA'_H_low},
\end{align}
where $\sigma_{\pm}=\left(I\pm\sigma_{z}\right)/2$ and $\hbar v_{F}=3\gamma_{0}a$/2, and as before $\mathbf{k}=(k_x,k_y)$ is a momentum measured relatively to the Dirac points.

\begin{figure}[h]
\centering{}\includegraphics[scale=0.95]{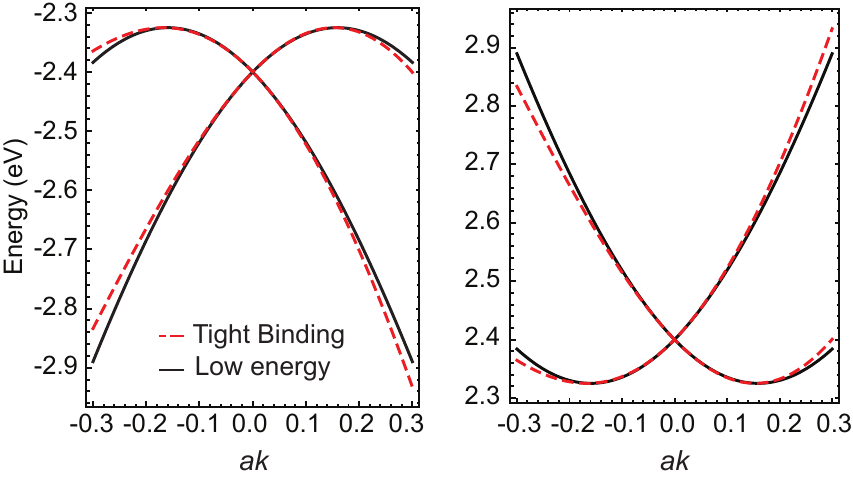}\caption{\label{fig:AA'_bands}Valence and conduction bands obtained from the
tight binding Hamiltonian, and the low energy approximation for the
AA' bilayer. The momenta is measured relatively to the K point (the
results near the K' point are identical). A good agreement is seen
between the two sets of data.}
\end{figure}

Diagonalizing this Hamiltonian, the following dispersion relation
is found:
\begin{align}
E_{\mathbf{k}}^{\lambda,\eta} & \approx\frac{\lambda}{2}\sqrt{E_{g}^{2}+\left(2\gamma_{1}+3\eta\gamma_{0}ak\right)^{2}}
\end{align}
with $\lambda=\pm1$ or $c/v$ depending if it is used as a number
or as an index, and $\eta=\pm1$. As in the case of the AB bilayer,
we see that the energy dispersion is independent of the valley index
$\tau$. The depiction of $E_{\mathbf{k}}^{\lambda,\eta}$ near the
$K$ point is given in Fig. \ref{fig:AA'_bands}. 
There, we see that the band structure of the AA' bilayer presents
a drastically different shape to that of the AB bilayer. While before
we found that the two valence/conduction bands were clearly separated
in energy, here wee see that a critical point exist at $k=0$ where
the bands touch. Moreover, contrarily to the AB bilayer, where the
extrema of the bands were located at zero momentum, here we find the
band maxima and minima at $k=2\gamma_{1}/3\gamma_{0}a$. The eigenvectors
found from the diagonalization of the low energy Hamiltonian are
\begin{align}
|u_{\mathbf{k}}^{v,\tau,\eta}\rangle & =\frac{1}{\sqrt{\mathcal{V}_{\eta,\mathbf{k}}}}\left[\begin{array}{c}
\frac{E_{g}-\sqrt{E_{g}^{2}+4\left(\gamma_{1}+\eta\hbar v_{F}k\right)^{2}}}{2\left(\gamma_{1}+\eta\hbar v_{F}k\right)}\\
\eta\tau e^{-i\tau\theta}\\
\eta\tau e^{-i\tau\theta}\frac{E_{g}-\sqrt{E_{g}^{2}+4\left(\gamma_{1}+\eta\hbar v_{F}k\right)^{2}}}{2\left(\gamma_{1}+\eta\hbar v_{F}k\right)}\\
1
\end{array}\right],\nonumber \\
|u_{\mathbf{k}}^{c,\tau,\eta}\rangle & =\frac{1}{\sqrt{\mathcal{C}_{\eta,\mathbf{k}}}}\left[\begin{array}{c}
\frac{E_{g}+\sqrt{E_{g}^{2}+4\left(\gamma_{1}+\eta\hbar v_{F}k\right)^{2}}}{2\left(\gamma_{1}+\eta\hbar v_{F}k\right)}\\
\eta\tau e^{-i\tau\theta}\\
\eta\tau e^{-i\tau\theta}\frac{E_{g}+\sqrt{E_{g}^{2}+4\left(\gamma_{1}+\eta\hbar v_{F}k\right)^{2}}}{2\left(\gamma_{1}+\eta\hbar v_{F}k\right)}\\
1
\end{array}\right].\label{eq:AA'_spinors}
\end{align}
where $\mathcal{C}_{\eta,\mathbf{k}}/\mathcal{V}_{\eta,\mathbf{k}}$
are normalization factors.

\subsection{Excitons and conductivity}

In order to obtain the excitonic energies and wave functions of the
AA' bilayer, one must return to the BSE, first presented in Eq. (\ref{eq:BSE}).
Because in the previous section we already discussed the nuances of
the BSE, and outlined our approach to solving it, we do not repeat
the same analysis here. Instead, we note only that the spinors given
in Eq. (\ref{eq:AA'_spinors}) already have the phase choice which
allows the transformation of the BSE from a 2D integral equation,
to a 1D problem (see Appendix \ref{sec:Solving-the-BSE} for details
on how to solve the 1D integral equation).

Considering a suspended bilayer $(\epsilon=1)$, and once again using
$r_{0}=16\text{Å}$ \citep{paleari2018excitons}, we solve the BSE
and find the energies and wave functions of the excitons for the AA'
bilayer. We stress that, similarly to the case of the AB configuration,
when the BSE was solved a corrected band gap of $E_{g}=6.96$ eV was
considered to match the value reported in Ref. \citep{paleari2018excitons}.
Once again, the exact value of the band gap should not have a significant
impact on the qualitative analysis of the results. As in the previous
section we solved the BSE using a 100-point Gauss-Legendre quadrature, which guaranteed the convergence of the excitonic energies and wave functions of the first ten states.

\begin{figure}[h]
\centering{}\includegraphics[scale=0.95]{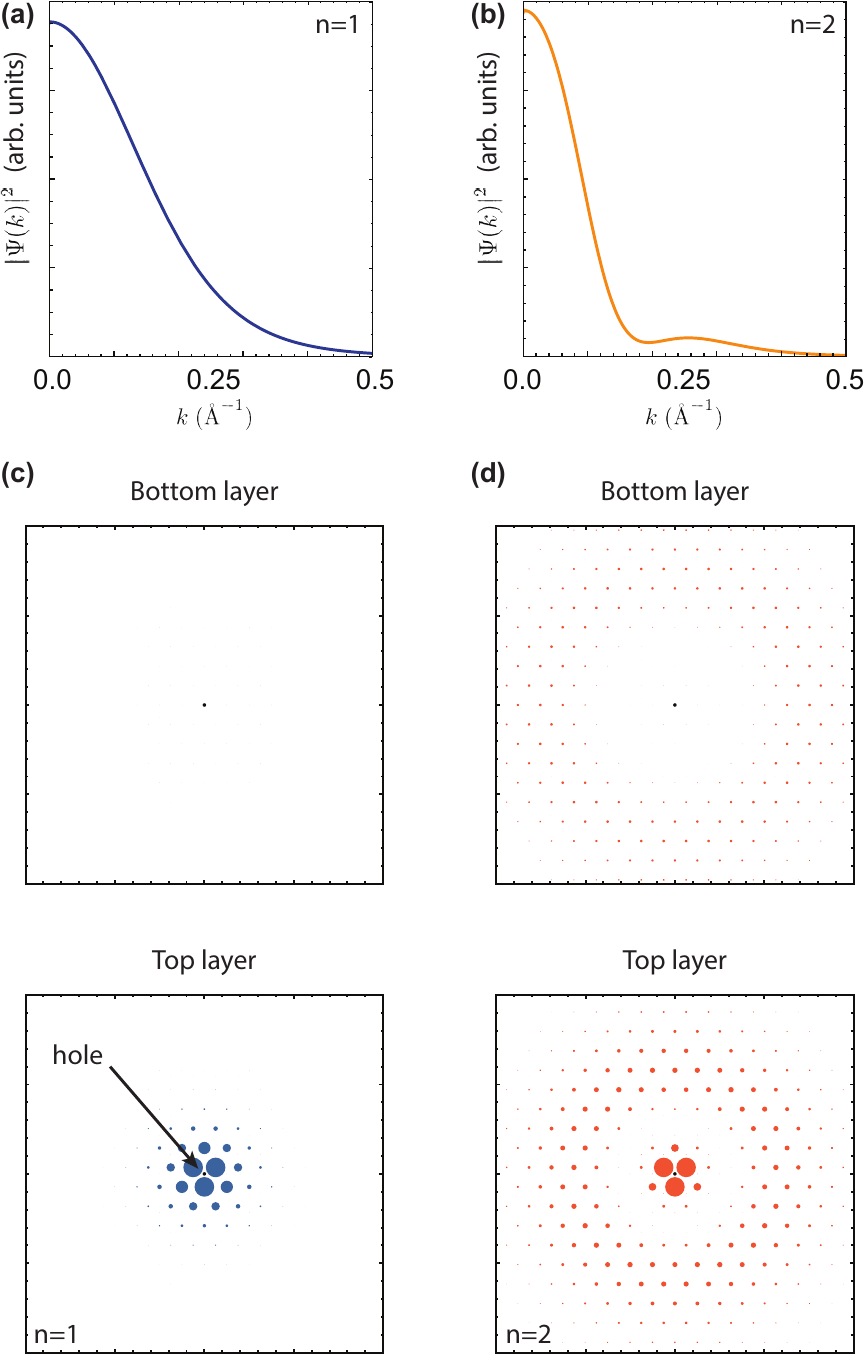}\caption{\label{fig:AAp_WF}(a)-(b) Wave functions, $|\boldsymbol{\Psi}(k)|^{2}=\sum_{cv}|\psi_{cv}(k)|^{2}$,
of the exciton states which are responsible for the first two resonances
of the optical conductivity (highlighted in the same color as the
respective wave function). (c)-(d) Real space representation
of the absolute value of the wave function when the hole is placed
in the $|t,2\rangle$ sub-lattice (containing a nitrogen atom). The position of the hole is marked by a black dot.}
\end{figure}

In Fig. \ref{fig:AAp_WF} (a) and (b) we depict the wave functions
$|\boldsymbol{\Psi}(k)|^{2}=\sum_{cv}|\psi_{cv}(k)|^{2}$ of two of
the states found by solving the BSE. 
These two states have energies
$E_{n=1}=5.64$ eV and $E_{n=2}=6.36$ eV and correspond to the first
two bright states of the system, i.e. the ones that originate the
first resonances of the optical conductivity (shown below). We note
that both of these states are doubly degenerate, without accounting
for spin or valley degeneracy. Analyzing the representation of the
wave functions in momentum space we realize that both resemble the
wave functions of the s-states of the Hydrogen atom, since both are
finite at the origin, and then decay to zero with zero and one nodes
for the $n=1$ and $n=2$ states, respectively. Because of the unique
band structure of the AA' bilayer, and contrarily to what we did in
the AB configuration, here we can't reduce the four band problem,
to an approximate two band one, since it is impossible to define a
pair of bands which could be considered the most relevant one for
the low energy response. Hence, the labeling of the these states as
s-states is based solely on their wave functions in analogy with the Hydrogen atom for which only s-states have finite wave functions at the origin.

In the panels (c) and (d) of Fig. \ref{fig:AAp_WF} we depict the
wave function of the two states in real space when the hole is placed
on the sub-lattice $|t,2\rangle$ (corresponding to a nitrogen atom), and find that both are mostly
intralayer excitons, since the real space wave function is mainly
distributed over same layer where the hole is located. If the hole is placed on the sub-lattice $|b,2\rangle$ (also a nitrogen atom) the results are identical to the ones depicted, only this time the wave function is almost entirely distributed over the bottom layer. When the hole is placed in either $|b/t,1\rangle$ sublattices (with boron atoms), the resulting real space wave function is essentially zero, indicating the preference of holes to appear on nitrogen atoms. These considerations are further backed by the values found when the wave function is integrated over each layer for a given position of the hole, which we show in Table \ref{tab: AAp_real_space_WF}.
\begin{table}[h]
\begin{tabular}{|l|l|l|l|l|l|}
\hline
                                             &        & $|1,b\rangle$ & $|2,b\rangle$ & $|1,t\rangle$ & $|2,t\rangle$ \\ \hline
\multicolumn{1}{|c|}{\multirow{2}{*}{$n=1$}} & Bottom & 0.01          & 0.46          & 0.00          & 0.02          \\ \cline{2-6} 
\multicolumn{1}{|c|}{}                       & Top    & 0.00          & 0.02          & 0.01          & 0.46          \\ \hline
\multirow{2}{*}{$n=2$}                       & Bottom & 0.01          & 0.33          & 0.00          & 0.14          \\ \cline{2-6} 
                                             & Top    & 0.00          & 0.14          & 0.01          & 0.33          \\ \hline
\end{tabular}
\caption{\label{tab: AAp_real_space_WF}Integrate values of the square modulus of the sublattice resolved real space wave function for the AA' bilayer. The horizontal top row indicated the sublattice where the hole is placed, while the two columns on the left indicate the exciton, and the layer where the electron is located.}
\end{table}
The identification of these
two states as being due to (mainly) intralayer s-excitons agrees with \citep{paleari2018excitons},
where the same conclusion was obtained from \emph{ab initio} calculations
and symmetry considerations. 

\begin{figure}[h]
\centering{}\includegraphics[scale=1.05]{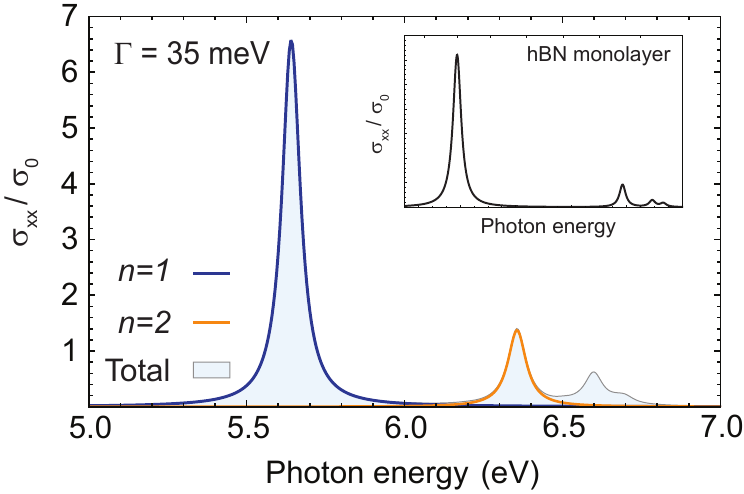}\caption{\label{fig:AAp_sigma_xx}Optical conductivity of an AA' hBN bilayer accounting for the first 10 exciton states with $m=\pm \tau$ (the contributions from both valleys were accounted for). The first two resonances correspond to
the excitonic states depicted in Fig. \ref{fig:AAp_sigma_xx}. To
obtain this result a phenomenological broadening of 35 meV was considered.
The conductivity is given in terms of the conductivity of graphene
$\sigma_{0}=e^{2}/4\hbar$. The inset shows a schematic depiction of the monolayer conductivity \citep{henriques2019optical}}
\end{figure}

Now that the solutions of the BSE were found, we can evaluate the longitudinal 
conductivity of the AA' bilayer. 
Using the definition given in Eq. (\ref{eq:Thomas_cond})
for the conductivity, we obtain the result depicted in Fig. \ref{fig:AAp_sigma_xx},
where the area shaded in blue corresponds to the conductivity obtained
accounting for 10 exciton states (once again states with $m=\pm\tau$ are selected, only this time both present identical oscillator strengths.). The dark blue and orange outlines
are the individual contributions of the two s-states whose wave functions
were depicted in Fig. \ref{fig:AAp_WF}.
First, we highlight the resemblance between our result and that of
Ref. \citep{paleari2018excitons}, especially for the first resonances,
where we see that the location of the first two peaks, as well as
their relative intensity, is similar in both works. At higher energies,
however, we observe significant differences between our conductivity
and the one obtained with \emph{ab initio} calculations. This mismatch
at higher energies was to be expected, since ours is a low energy
theory, incapable of capturing the more nuanced features near the
band edge. Nonetheless, the similarities at lower energies are a good
indicator of the validity of our results. The conductivity of the AA' bilayer resembles that of the monolayer \citep{henriques2019optical}, since in both cases the s-states are the bright one, and both present a set of resonances with monotically decreasing oscillator strength. When compared to the conductivity of the AB bilayer, we find that the small peak between the first two resonances of Fig.
\ref{fig:AAp_sigma_xx} is absent in the AA' configuration; hence, this small resonance can then be seen as a fingerprint of the AB stacking. 

\subsection{The effect of bias}

One of the main features of the AA' bilayer is its peculiar band structure,
particularly the degeneracy at $k=0$. An interesting thing to consider
is the effect of lifting said degeneracy. To study this possibility,
we now briefly consider the case of a biased AA' bilayer. The bias
can be introduced in the system trough the application of a vertical
displacement field. Since the application of such a field breaks the
inversion symmetry of the AA' bilayer, one may expect new optical
selection rules for the biased bilayer when compared to the unbiased
case.

To introduce the effect of bias in our low energy model, we need only
add a new contribution to the low energy Hamiltonian given in Eq.
(\ref{eq:AA'_H_low}):
\begin{equation}
H_{\textrm{bias}}=VI\otimes\sigma_{z},
\end{equation}
with $V$ the quantifying the magnitude of the bias, $I$ the identity
matrix and $\sigma_{z}$ the $z$ Pauli matrix. 
\begin{figure}[h]
\centering{}\includegraphics[scale=1.1]{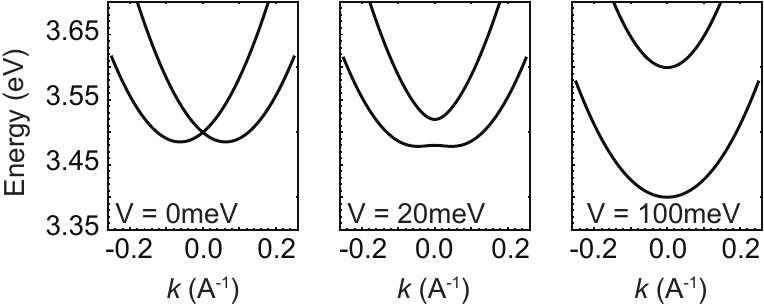}\caption{\label{fig:Bias_bands}Conduction bands of the biased AA' bilayer,
for three different bias values, $V=0$ meV, $V=20$ meV and $V=100$
meV. The valence bands present and identical dispersion relation.}
\end{figure}
The bands associated with this new Hamiltonian are depicted in Fig.
\ref{fig:Bias_bands}, where we see that for a small bias the degeneracy
at $k=0$ is indeed lifted, and the lower energy conduction band acquires
the form of a Mexican hat, similar to what is found in biased bilayer
graphene. As the bias increases, so does the separation between the
two bands, and the shape of the bottom band becomes closer to a simple
parabolic dispersion. We also note that, although we only show the
results for positive bias, the bands for negative bias are identical
to the ones presented here. 

\begin{figure}[h]
\centering{}\includegraphics{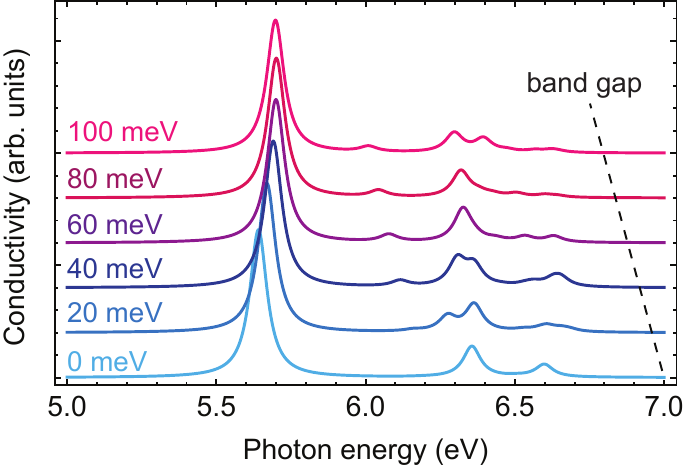}\caption{\label{fig:Bias_cond}Optical conductivity for the biased AA' bilayer
for different bias values. The results for negative bias are identical.
The different conductivities are vertically shifted for clarity. Also
depicted is the location of the band gap for each bias value.}
\end{figure}

Solving the BSE (using same parameters
we used in the unbiased case) with this new Hamiltonian, and computing
the conductivity due to an in-plane linearly polarized external electric
field, we obtain the result depicted in Fig. \ref{fig:Bias_cond},
where different values for $V$ are considered (the results for negative
bias are identical).
First, we see that, as expected, for zero bias the result of Fig.
\ref{fig:AAp_sigma_xx} is recovered. Then, as the bias increases
we observe that the first resonance is shifted to higher energies
despite the reduction of the band gap, reflecting a reduction of its
binding energy. Furthermore, we note that the initially simple features
at higher energies become significantly more complex as the bias increases,
since different new small resonances start to appear. From the inspection
of the real space wave functions, we find that these new small resonances
are associated with excitons which appear to be mainly interlayer
(electron and hole in opposite layers). The excitons originating the
larger resonances, which in the unbiased case were almost entirely
intralayer, see their interlayer component increase. At last we note that in the presence of external bias the conductivity becomes more alike the one found for the AB bilayer, something we assign to the breaking of inversion symmetry.

As we saw in Fig. \ref{fig:AAp_WF}, at zero bias, the resonances
on the conductivity are essentially due to s-states, whose wave functions
are finite at $k=0$. However, as soon as some bias is introduced
in the system, the symmetry of the problem changes, and the optical
selection rules are affected. Since for the biased bilayer we can
split the bands into low energy and high energy groups, an approximate
two band model can be employed (see Sec. \ref{subsec:Complementary-analysis}),
allowing us to establish approximate optical selection rules. Applying
the Lowdin partitioning, as described prior to Eq. (\ref{eq:Lowdin}),
we find the following effective two band Hamiltonian:
\begin{equation}
H_{\textrm{Bias, eff}}=\left(\begin{array}{cc}
\Delta(-|V|) & -\frac{9\gamma_{0}^{2}\gamma_{1}\Delta(|V|)}{8V^{3}-2VE_{g}^{2}}k^{2}\\
-\frac{9\gamma_{0}^{2}\gamma_{1}\Delta(|V|)}{8V^{3}-2VE_{g}^{2}}k^{2} & -\Delta(-|V|)
\end{array}\right),
\end{equation}
where $\Delta(x)=\sqrt{\left(E_{g}/2+x\right)^{2}+\gamma_{1}^{2}}.$
In its current form, this Hamiltonian holds for both $V>0$ and $V<0$
(but clearly fails to describe $V=0$, (the unbiased case). Using
the procedure of Ref. \citep{Zhang2018} one more time, we identify
the winding number as $w=0$, and as a consequence the bright excitons
are those with $m_X=\pm\tau$, that is, p-states. If trigonal warping
had been considered, for example by including non-vertical interlayer
hoppings, then states with $m=\pm2$ and $m=\pm4$ could also be excited (due to an additional factor of 3 stemming from the lattice symmetry contributing to $m_X$).
Note how for the biased AA' bilayer the s-states are dark even if
trigonal warping is considered, in stark contrast with the unbiased
case, where s-states dominate the optical response.

\section{Discussion\label{sec:Discussion}}

In this paper we studied the optical conductivity due to excitonic
effects of two types of hBN bilayers, the AB and AA' configurations.
The comprehension of the properties of these bilayers is of great
utility in the study of twisted bilayers at arbitrary angles, since
the results we presented correspond to the limit cases of $0^\circ$ and $60^\circ$
rotation.

To obtain the excitonic spectrum of each type of bilayer we solved
the Bethe-Salpeter equation (BSE) using the Bloch factors given by
a low energy four-band Hamiltonian. To ease the numerical weight of the calculation
we avoided the process of solving a 2D integral equation by a judicious 
choice of the phases of the Bloch factors, allowing us to cast the
BSE into a 1D problem, which can then be solved in a rather efficient way. We emphasize that the method we presented to solve the four-band BSE gives better results than those of effective theories, such as the
Lowdin partitioning. Although useful to extract optical selection
rules, this type of effective approach fails to accurately predict the optical response (as we saw for the AB bilayer), and may even be impossible to apply (as we saw for the AA' bilayer).
Moreover, our approach is far less computationally expensive
than methods which require the solution of the BSE in two dimensions, allowing the exploration of such systems by a broader audience.

Regarding the conductivities of the two considered bilayers, we found that the AB configuration presents an optical response where both intralayer and interlayer excitons participate. In particular, we found the first (and largest) excitonic resonance to be due to a mainly intralayer exciton, followed by a small, yet well resolved, resonance due to an interlayer exciton (which is only captured when the four bands of the model are accounted for); this small peak is followed by a larger one, also due to a mainly interlayer exciton. Furthermore, we found that for the AB bilayer the two main resonances in the optical conductivity could be assigned with the Hydrogenic label of p-states (angular
quantum number equal to 1); f-states (angular quantum number 3) are also allowed to be excited albeit with tiny oscillator strengths, and s-states
(angular quantum number 0) appear if trigonal warping is accounted for.

For the AA' bilayer we found an optical conductivity dominated by mainly intralayer excitons, to which we assigned the Hydrogenic label of s-states due to the lineshape of the wave functions in momentum space, in agreement with \citep{Alejandro2016}. Contrarily to the AB stacking, the conductivity of the AA' bilayer presented a set of resonances with monotonically decreasing magnitude (similar to the monolayer). Hence, the small peak between two larger ones in the AB bilayer is a clear differentiating feature between the two considered stakings.

When the case of a biased AA' bilayer was studied, we found that the s-states became dark, and the p-states dominated the optical spectrum; the change of optical selection rules is a consequence of the symmetry breaking introduced by the bias. Moreover, as the bias increased we found that the first (and more pronounced) resonance was shifted to higher energies, going against the trend of the band gap, which decreased with increasing bias. We also found that the introduction of the bias lead to an overall more complex optical response, due to the increased contribution from interlayer excitons.


\begin{acknowledgments}
B.A., R.M.R and N.M.R.P acknowledge support by the Portuguese Foundation for Science and Technology (FCT) in the framework of the Strategic Funding UIDB/04650/2020. J.C.G.H. acknowledges the Center of Physics for a grant funded by the UIDB/04650/2020 strategic project. B.A. and N.M.R.P acknowledge support from FCT-Portugal through Project EXPL/FIS-MAC/0953/2021. B.A. further acknowledges funding from FCT-Portugal via Grant CEECIND/02936/2017. R.M.R. and N.M.R.P. also acknowledge support from the European Commission through the project GrapheneDriven Revolutions in ICT and Beyond (Ref. No. 881603, CORE 3) and the project PTDC/FIS-MAC/2045/2021. N.M.R.P. further acknowledges COMPETE 2020, PORTUGAL 2020, FEDER and the FCT through projects POCI-01-0145-FEDER-028114, POCI-01-0145-FEDER-02888 and PTDC/NANOPT/29265/2017.
\end{acknowledgments}

\appendix

\section{Details on the DFT calculations\label{sec:Details-on-DFT}}

\begin{figure}[h]
\centering{}\includegraphics[scale=0.8]{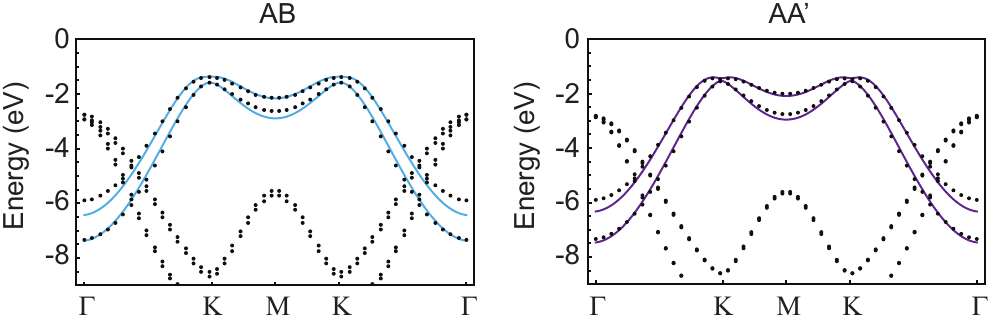}\caption{\label{fig:Fit-DFT}Fit of the tight binding bands to the results
found from DFT calculations. For the AB bilayer we find $E_{g}=4.585$
eV, $\gamma_{0}=2.502$ eV and $\gamma_{1}=0.892$ eV. For the AA'
bilayer we obtain $E_{g}=4.650$ eV, $\gamma_{0}=2.491$ eV and $\gamma_{1}=0.595$
eV.}
\end{figure}

Density Functional Theory (DFT) calculations were performed using
the software package {\sc Quantum ESPRESSO} \citep{QE-2009}. We
used a scalar-relativistic norm-conserving pseudopotential \citep{Hamman2013,schlipf2015optimization}
and the generalized gradient approximation of Perdew-Burke-Ernzerhof
(GGA-PBE) \citep{PhysRevLett.77.3865}. The plane-wave cut-off was
80 Ry and for the integration over the Brillouin-zone the scheme proposed
by Monkhorst-Pack \citep{PhysRevB.13.5188} with a grid of $18\times18\times1$
$\mathbf{k}$-points was used. A vacuum size between the layer images
of 25 bohr was enough to avoid interactions between the periodic images.
We also included the van der Waals correction proposed by Grimme.\citep{grimme2006semiempirical,barone2009role}.
Atoms were relaxed to establish the spacing between layers. The tight-binding
parameters were obtained by fitting the DFT bands along a path in
the first Brillouin zone as depicted in Fig. \ref{fig:Fit-DFT}. Only
the valence bands were fitted, since the DFT calculations capture
less accurately the empty states of the conduction bands \citep{Ribeiro2011}.

\section{On the exciton's angular quantum number\label{sec:On-the-exciton's}}

\subsection{Two band system}

Let us start by considering the problem of an hBN monolayer, which
we take as a concrete example of a two band system \citep{henriques2019optical}.
We model the monolayer with a two band Dirac Hamiltonian to describe
its low energy electronic properties. From the diagonalization of
the Hamiltonian, one easily shows that the Bloch factors take the
form \citep{Pedersen2019}:
\begin{align}
|u_{c,\mathbf{k}}\rangle & =\left[e^{i\theta}\sin\xi_{k},\cos\xi_{k}\right]^{\textrm{T}},\\
|u_{v,\mathbf{k}}\rangle & =\left[e^{i\theta}\cos\xi_{k},-\sin\xi_{k}\right]^{\textrm{T}},
\end{align}
where $c/v$ labels the conduction/valence band, $\theta=\arctan k_{y}/k_{x}$
and $\xi_{k}$ is a function which approaches zero as the momentum
$k$ vanishes. Alternatively, we could have defined the Bloch factors
as: 
\begin{align}
|w_{c,\mathbf{k}}\rangle & =\left[e^{i\theta}\sin\xi_{k},\cos\xi_{k}\right]^{\textrm{T}},\\
|w_{v,\mathbf{k}}\rangle & =\left[\cos\xi_{k},-e^{-i\theta}\sin\xi_{k}\right]^{\textrm{T}},
\end{align}
since state vectors are only defined up to a global phase factor.

Let us now introduce excitons in this system. We consider that, as
in the main text, the wave function of an exciton in momentum space
can be written as $\psi_{cv}(\mathbf{k})=f_{cv}(k)e^{im\theta}$. The real space wave function can the be defined as
\begin{align}
	\Psi_{\alpha \beta}^u (\mathbf{r}_e,\mathbf{r}_h) = \sum_\mathbf{k} e^{i(\boldsymbol{K}+\mathbf{k})\cdot(\mathbf{r}_e - \mathbf{r}_h)} f_{cv}(k) e^{im_u\theta} u^\alpha_{\mathbf{k},c} \left( u^\beta_{\mathbf{k},v} \right)^*  \\
	\Psi_{\alpha \beta}^w (\mathbf{r}_e,\mathbf{r}_h) = \sum_\mathbf{k} e^{i(\boldsymbol{K}+\mathbf{k})\cdot(\mathbf{r}_e - \mathbf{r}_h)} f_{cv}(k) e^{im_w\theta} w^\alpha_{\mathbf{k},c} \left( w^\beta_{\mathbf{k},v} \right)^*
\end{align}
where $\mathbf{r}_{e}$ and $\mathbf{r}_{h}$ are the electron and
hole positions, respectively, and $u/w_{\mathbf{k},c}^{\alpha}$ refers
to the $\alpha$ sub-lattice entry of the Bloch factor $|u/w_{\mathbf{k},c}\rangle$
(an analogous definition holds for $u/w_{\mathbf{k},v}^{\beta}$)
From the definition of
the Bloch factors, and recalling that $\lim_{k\rightarrow0}\xi_{k}=0$,
we see that the product $u^\alpha_{\mathbf{k},c} \left( u^\beta_{\mathbf{k},v} \right)^*$
approximately introduces a phase $e^{-i\theta}$ in the definition
of the wave function, while $w^\alpha_{\mathbf{k},c} \left( w^\beta_{\mathbf{k},v} \right)^*$ introduces
no phase. Hence, when we define the wave function
with the $u$-Bloch factors, we find a pseudo-spin angular quantum
number of $m_{u}^{\textrm{ps}}=-1$, while for the $w$-Bloch factors
we have $m_{w}^{\textrm{ps}}=0$. Notice how we focused our analysis
near $k=0$, since that is where selection rules are stronger; momentum
dependence tens to weaken optical selection rules.
%

For a linearly polarized electric field, one can show that the optical
response is proportional to $\Omega^x_{u/w}=\sum_{\mathbf{k}}\psi_{cv}(\mathbf{k})\langle u/w_{v,\mathbf{k}}|x|u/w_{c,\mathbf{k}}\rangle$
\citep{Cao2018,Zhang2018}. Like we did in the main text, the matrix
element of the position operator can be found from the commutator
of the Hamiltonian with the position operator itself. As we said in
the beginning, we are considering a Dirac Hamiltonian to model the
system. Because of that, we can write $\Omega_{u/w}$ as
\begin{align}
\Omega^x_{u} & \propto\sum_{\mathbf{k}}f_{cv}(k)e^{im_{u}\theta}\frac{\langle u{}_{v,\mathbf{k}}|\sigma_{x}|u{}_{c,\mathbf{k}}\rangle}{E_{v,k}-E_{c,k}}\\
\Omega^x_{w} & \propto\sum_{\mathbf{k}}f_{cv}(k)e^{im_{w}\theta}\frac{\langle w{}_{v,\mathbf{k}}|\sigma_{x}|w{}_{c,\mathbf{k}}\rangle}{E_{v,k}-E_{c,k}},
\end{align}
with $E_{v/c,k}$ the dispersion relations of the model Hamiltonian,
which are obviously independent of the phase choice for the Bloch
factors. Converting the sum over $\mathbf{k}$ into a 2D integral
in momentum space, and carrying out the necessary calculations, one
finds that $\Omega_{u}$ and $\Omega_{w}$ are only finite if $m_{u}=\pm1$
and $m_{w}=0,-2$, respectively. Thus, at first, it may appear that
the choice of phase for the Bloch factors changes the optical selection
rules, since different angular dependencies for the exciton envelope
function are selected. However, when the contribution of the pseudo-spin
angular quantum number is taken into account, we see that $m_{u}+m_{u}^{\textrm{ps}}=m_{w}+m_{w}^{\textrm{ps}}=0,-2$.
The sum of these two contributions is independent of the phase chosen
for the Bloch factors, and is the appropriate angular quantum number
\citep{park2010tunable}.

\subsection{Four band system}

In the first part of this appendix we saw how to define the appropriate
angular quantum number for a two band system such as an hBN monolayer.
To achieve this one must sum the angular quantum number from the excitonic
envelope function with the angular quantum number given by the Bloch
factors, to obtain the appropriate angular quantum number; while the
first two depend on the phase chosen for the Bloch factors, the last
one is independent of it (as it should, in order to be an approximately
good quantum number).

Let us now consider a four band model, such as the ones treated in
main text. For such a system, the real space exciton wave function reads
\begin{align}
	\Psi_{\alpha,\beta} (\mathbf{r}_e,\mathbf{r}_h) = \sum_{\mathbf{k},c,v} e^{i(\boldsymbol{K}+\mathbf{k})\cdot(\mathbf{r}_e - \mathbf{r}_h)} f_{cv}(k) e^{im\theta} u^\alpha_{\mathbf{k},c} \left( u^\beta_{\mathbf{k},v} \right)^*
\end{align}

which differs from the definition given in the first part of this appendix due to the sums over the bands.
The
problem in defining an angular quantum number for the exciton in a
four band system lies in the definition of the pseudo-spin contribution.
While the contribution from the envelope function to the angular quantum
number is still well defined, the same can not be said for the the
pseudo-spin part, since, in principle, each of the terms $|u_{c,\mathbf{k}}\rangle\langle u_{v,\mathbf{k}}|$
can contribute with a different complex exponential (which is the
case for the two systems treated in the main text), thus stopping
us from obtaining a well defined $m_{\textrm{ps}}$, with which the
appropriate angular quantum number of the exciton (independent of
phase choices) could be determined. Although this could be bypassed
with a phase choice that, for example, left all the spinors without
complex exponentials in the $k\rightarrow0$ limit, that would no
be helpful for our approach, where a specific phase choice has to
be performed to cast the BSE into a 1D problem, thus simplifying its numerical solution.

\begin{widetext}

\section{\label{sec:Solving-the-BSE}Solving the BSE}

In this appendix we shall give a more in depth description on how
to numerically solve the Bethe-Salpeter equation (BSE) presented in
the main text. The method we present is an extension of the one applied
for the Hydrogen atom in Ref. \citep{Chao1991}. We take Eq. (\ref{eq:simpler_BSE})
of the main text as our starting point:
\begin{equation}
\left(E_{k}^{c}-E_{k}^{v}\right)f_{cv}(k)-\sum_{c'v'}\int qdqd\theta_{q}V(\mathbf{k}-\mathbf{q})\langle u_{\mathbf{k}}^{c}|u_{\mathbf{q}}^{c'}\rangle\langle u_{\mathbf{q}}^{v'}|u_{\mathbf{k}}^{v}\rangle f_{c'v'}(q)e^{im\left(\theta_{q}-\theta_{k}\right)}=Ef_{cv}(k).
\end{equation}
As discussed in the main text, we consider the spinor product to have
the following form
\begin{equation}
\langle u_{\mathbf{k}}^{c}|u_{\mathbf{q}}^{c'}\rangle\langle u_{\mathbf{q}}^{v'}|u_{\mathbf{k}}^{v}\rangle=\sum_{\lambda}\mathcal{A}_{\lambda}^{cc'vv'}(k,q)e^{i\lambda(\theta_{q}-\theta_{k})},
\end{equation}
where $\lambda$ is some integer, and $\mathcal{A}_{\lambda}^{cc'vv'}(k,q)$
are coefficients determined by the explicit computation of the spinor
product. Inserting this into the previous equation, and noting that
$V(\mathbf{k}-\mathbf{q})\equiv V(k,q,\theta_{q}-\theta_{k})$, one
finds
\begin{equation}
\left(E_{k}^{c}-E_{k}^{v}\right)f_{cv}(k)-\sum_{c'v'}\sum_{\lambda}\int qdqd\vartheta V(k,q,\vartheta)\mathcal{A}_{\lambda}^{cc'vv'}(k,q)f_{c'v'}(q)e^{i\left(m+\lambda\right)\vartheta}=Ef_{cv}(k),
\end{equation}
where we introduced the variable change $d\theta_{q}\rightarrow d\vartheta$
with $\vartheta=\theta_{q}-\theta_{k}$. Now, recalling the definition
of $V(k,q,\vartheta)$, we introduce a new function, $\mathcal{I}_{\nu}(k,q)$,
which corresponds to the integral over $d\vartheta$, that is
\begin{equation}
\mathcal{I}_{\nu}(k,q)=\int_{0}^{2\pi}\frac{\cos\left(\nu\vartheta\right)}{\kappa(k,q,\vartheta)\left[1+r_{0}\kappa(k,q,\vartheta)\right]}d\vartheta
\end{equation}
with $\kappa(k,q,\vartheta)=\sqrt{k^{2}+q^{2}-2kq\cos\vartheta}$.
Notice how only $\cos(\nu\theta)$ enters the integral, since the
analogous term in $\sin(\nu\vartheta)$ vanishes by symmetry.From
inspection, it should be clear that when $q=k$ the function $\mathcal{I}_{\nu}(k,q)$
is numerically ill-behaved, and as such must be treated carefully.
Looking at its definition, one sees that we can express $\mathcal{I}_{\nu}(k,q)$
in terms of partial fractions as
\begin{align}
\mathcal{I}_{\nu}(k,q) & =\int_{0}^{2\pi}\frac{\cos\left(\nu\vartheta\right)}{\kappa(k,q,\vartheta)}d\vartheta-r_{0}\int_{0}^{2\pi}\frac{\cos\left(\nu\vartheta\right)}{\left[1+r_{0}\kappa(k,q,\vartheta)\right]}d\vartheta\\
 & \equiv\mathcal{J}_{\nu}(k,q)-\mathcal{K}_{\nu}(k,q),
\end{align}
where from these two terms only the first one, $\mathcal{J}_{\nu}(k,q)$,
is problematic when $k=q$, since $\mathcal{K}_{\nu}(k,q)$ contains
an additional 1 in the denominator which prevents any divergence.
Before we explain how to avoid this numerical problem, let us first
express the BSE in a more convenient manner. First, we write
\begin{equation}
\left(E_{k}^{c}-E_{k}^{v}\right)f_{cv}(k)-\sum_{c'v'}\sum_{\lambda}\int_{0}^{\infty}\left\{ \mathcal{J}_{m+\lambda}(k,q)\mathcal{A}_{\lambda}^{cc'vv'}(k,q)f_{c'v'}(q)-\mathcal{K}_{m+\lambda}(k,q)\mathcal{A}_{\lambda}^{cc'vv'}(k,q)f_{c'v'}(q)\right\} qdq=Ef_{cv}(k).
\end{equation}
Then, we define $\mathcal{B}_{m}^{cc'vv'}(k,q)=\sum_{\lambda}\mathcal{J}_{m+\lambda}(k,q)\mathcal{A}_{\lambda}^{cc'vv'}(k,q)$
and $\mathcal{C}_{m}^{cc'vv'}(k,q)=\sum_{\lambda}\mathcal{K}_{m+\lambda}(k,q)\mathcal{A}_{\lambda}^{cc'vv'}(k,q)$.
With these new definitions, one finds
\begin{equation}
\left(E_{k}^{c}-E_{k}^{v}\right)f_{cv}(k)-\sum_{c'v'}\int_{0}^{\infty}\mathcal{B}_{m}^{cc'vv'}(k,q)f_{c'v'}(q)qdq+\sum_{c'v'}\int_{0}^{\infty}\mathcal{C}_{m}^{cc'vv'}(k,q)f_{c'v'}(q)qdq=Ef_{cv}(k).
\end{equation}

Now, let us focus on the numerical problem associated with $\mathcal{B}_{m}^{cc'vv'}(k,q)$.
To treat the divergence that appears when $k=q$, we introduce an
auxiliary function $g_{m}(k,q)$ and introduce the modification
\begin{equation}
\int_{0}^{\infty}\mathcal{B}_{m}^{cc'vv'}(k,q)f_{c'v'}(q)qdq\rightarrow\int_{0}^{\infty}\left[\mathcal{B}_{m}^{cc'vv'}(k,q)f_{c'v'}(q)-g_{m}(k,q)f_{c'v'}(k)\right]qdq+f_{c'v'}(k)\int_{0}^{\infty}g_{m}(k,q)qdq,
\end{equation}
with $g_{m}$ defined in such a way that $\lim_{q\rightarrow k}\left[\mathcal{B}_{m}^{cc'vv'}(k,q)-g_{m}(k,q)\right]=0$.
Following Ref. \citep{Chao1991}, we define $g_{m}$ as
\begin{equation}
g_{m}=\mathcal{B}_{m}^{cc'vv'}(k,q)\frac{2k^{2}}{k^{2}+q^{2}}
\end{equation}

With the analytical part of the calculation taken care of, we shall
now discuss how to numerically solve the equation we have arrived
to. To achieve this, we first introduce a variable change which transforms
the improper integral over $[0,\infty)$, into one with finite integration
limits, such as $[0,1]$; with this goal in mind we introduce $q=\tan[\pi x/2]$.
Afterwards, we discretize the variables $k$ and $x$ (and consequently
$q$), and find
\begin{align}
 & \left(E_{k_{i}}^{c}-E_{k_{i}}^{v}\right)f_{cv}(k_{i})+\sum_{c'v'}\sum_{j=1}^{N}\mathcal{C}_{m}^{cc'vv'}(k_{i},q_{j})f_{c'v'}(q_{j})q_{j}\frac{dq}{dx_{j}}\nonumber \\
 & -\sum_{c'v'}\sum_{j\neq i}\mathcal{B}_{m}^{cc'vv'}(k_{i},q_{j})f_{c'v'}(q_{j})q_{j}\frac{dq}{dx_{j}}w_{j}-f_{c'v'}(k_{i})\left\{ \int_{0}^{\infty}g_{m}(k_{i},p)pdp-\sum_{j\neq i}g_{m}(k_{i},q_{j})q_{j}\frac{dq}{dx_{j}}\right\} =Ef_{cv}(k_{i})
\end{align}
where $N$ is the number of points and $w_{j}$ is the weight function
of the chosen numerical quadrature; also, $q_{j}\equiv q(x_{j})$
and $dq/dx_{j}\equiv\left[dq/dx\right]_{x=x_{j}}$. Furthermore, we
note that $\int_{0}^{\infty}g_{m}(k_{i},p)pdp$ is numerically well
behaved as opposed to the original integral, $\int_{0}^{\infty}\mathcal{B}_{m}^{cc'vv'}(k_{i},p)pdp$.
Regarding the choice of quadrature, we employ a Gauss-Legendre quadrature,
which is defined as \citep{kythe2011computational}
\begin{equation}
\int_{a}^{b}f(y)dy\approx\sum_{i=1}^{N}w_{i}f(y_{i})
\end{equation}
where 
\begin{equation}
y_{i}=\frac{a+b+(b-a)\xi_{i}}{2},
\end{equation}
with $\xi_{i}$ the $i-$th zero of the Legendre polynomial $P_{N}(y)$,
and
\begin{equation}
w_{i}=\frac{b-a}{(1-\xi_{i})^{2}\left[P'_{N}(\xi_{i})\right]^{2}}
\end{equation}
with $P'_{N}(\xi_{i})\equiv\left[dP_{N}(y)/dy\right]_{y=\xi_{i}}$.

At last, the only thing left to do is to realize that this equation
can be expressed as an eigenvalue problem of a $4N\times4N$ matrix.
This matrix can be thought of as a $4\times4$ matrix of matrices,
each one with dimensions $N\times N$. The 16 blocks come from the
different combinations of the indexes $c$, $c'$, $v$ and $v'$,
with each block corresponding to a $N\times N$ matrix stemming from
the numerical discretization of the integral. Solving the eigenvalue
problem one finds the exciton energies and wave functions.\end{widetext}\bibliographystyle{aapmrev4-2}

\begin{thebibliography}{50}%
\makeatletter
\providecommand \@ifxundefined [1]{%
 \@ifx{#1\undefined}
}%
\providecommand \@ifnum [1]{%
 \ifnum #1\expandafter \@firstoftwo
 \else \expandafter \@secondoftwo
 \fi
}%
\providecommand \@ifx [1]{%
 \ifx #1\expandafter \@firstoftwo
 \else \expandafter \@secondoftwo
 \fi
}%
\providecommand \natexlab [1]{#1}%
\providecommand \enquote  [1]{``#1''}%
\providecommand \bibnamefont  [1]{#1}%
\providecommand \bibfnamefont [1]{#1}%
\providecommand \citenamefont [1]{#1}%
\providecommand \href@noop [0]{\@secondoftwo}%
\providecommand \href [0]{\begingroup \@sanitize@url \@href}%
\providecommand \@href[1]{\@@startlink{#1}\@@href}%
\providecommand \@@href[1]{\endgroup#1\@@endlink}%
\providecommand \@sanitize@url [0]{\catcode `\\12\catcode `\$12\catcode
  `\&12\catcode `\#12\catcode `\^12\catcode `\_12\catcode `\%12\relax}%
\providecommand \@@startlink[1]{}%
\providecommand \@@endlink[0]{}%
\providecommand \url  [0]{\begingroup\@sanitize@url \@url }%
\providecommand \@url [1]{\endgroup\@href {#1}{\urlprefix }}%
\providecommand \urlprefix  [0]{URL }%
\providecommand \Eprint [0]{\href }%
\providecommand \doibase [0]{https://doi.org/}%
\providecommand \selectlanguage [0]{\@gobble}%
\providecommand \bibinfo  [0]{\@secondoftwo}%
\providecommand \bibfield  [0]{\@secondoftwo}%
\providecommand \translation [1]{[#1]}%
\providecommand \BibitemOpen [0]{}%
\providecommand \bibitemStop [0]{}%
\providecommand \bibitemNoStop [0]{.\EOS\space}%
\providecommand \EOS [0]{\spacefactor3000\relax}%
\providecommand \BibitemShut  [1]{\csname bibitem#1\endcsname}%
\let\auto@bib@innerbib\@empty
\bibitem [{\citenamefont {Caldwell}\ \emph {et~al.}(2019)\citenamefont
  {Caldwell}, \citenamefont {Aharonovich}, \citenamefont {Cassabois},
  \citenamefont {Edgar}, \citenamefont {Gil},\ and\ \citenamefont
  {Basov}}]{caldwell2019photonics}%
  \BibitemOpen
  \bibfield  {author} {\bibinfo {author} {\bibfnamefont {J.~D.}\ \bibnamefont
  {Caldwell}}, \bibinfo {author} {\bibfnamefont {I.}~\bibnamefont
  {Aharonovich}}, \bibinfo {author} {\bibfnamefont {G.}~\bibnamefont
  {Cassabois}}, \bibinfo {author} {\bibfnamefont {J.~H.}\ \bibnamefont
  {Edgar}}, \bibinfo {author} {\bibfnamefont {B.}~\bibnamefont {Gil}},\ and\
  \bibinfo {author} {\bibfnamefont {D.}~\bibnamefont {Basov}},\ }\href@noop {}
  {\bibfield  {journal} {\bibinfo  {journal} {Nature Reviews Materials}\
  }\textbf {\bibinfo {volume} {4}},\ \bibinfo {pages} {552} (\bibinfo {year}
  {2019})}\BibitemShut {NoStop}%
\bibitem [{\citenamefont {Wang}\ \emph {et~al.}(2018)\citenamefont {Wang},
  \citenamefont {Chernikov}, \citenamefont {Glazov}, \citenamefont {Heinz},
  \citenamefont {Marie}, \citenamefont {Amand},\ and\ \citenamefont
  {Urbaszek}}]{Wang2018Colloquium}%
  \BibitemOpen
  \bibfield  {author} {\bibinfo {author} {\bibfnamefont {G.}~\bibnamefont
  {Wang}}, \bibinfo {author} {\bibfnamefont {A.}~\bibnamefont {Chernikov}},
  \bibinfo {author} {\bibfnamefont {M.~M.}\ \bibnamefont {Glazov}}, \bibinfo
  {author} {\bibfnamefont {T.~F.}\ \bibnamefont {Heinz}}, \bibinfo {author}
  {\bibfnamefont {X.}~\bibnamefont {Marie}}, \bibinfo {author} {\bibfnamefont
  {T.}~\bibnamefont {Amand}},\ and\ \bibinfo {author} {\bibfnamefont
  {B.}~\bibnamefont {Urbaszek}},\ }\href
  {https://doi.org/10.1103/RevModPhys.90.021001} {\bibfield  {journal}
  {\bibinfo  {journal} {Rev. Mod. Phys.}\ }\textbf {\bibinfo {volume} {90}},\
  \bibinfo {pages} {021001} (\bibinfo {year} {2018})}\BibitemShut {NoStop}%
\bibitem [{\citenamefont {Dean}\ \emph {et~al.}(2010)\citenamefont {Dean},
  \citenamefont {Young}, \citenamefont {Meric}, \citenamefont {Lee},
  \citenamefont {Wang}, \citenamefont {Sorgenfrei}, \citenamefont {Watanabe},
  \citenamefont {Taniguchi}, \citenamefont {Kim}, \citenamefont {Shepard},\
  and\ \citenamefont {Hone}}]{dean2010boron}%
  \BibitemOpen
  \bibfield  {author} {\bibinfo {author} {\bibfnamefont {C.~R.}\ \bibnamefont
  {Dean}}, \bibinfo {author} {\bibfnamefont {A.~F.}\ \bibnamefont {Young}},
  \bibinfo {author} {\bibfnamefont {I.}~\bibnamefont {Meric}}, \bibinfo
  {author} {\bibfnamefont {C.}~\bibnamefont {Lee}}, \bibinfo {author}
  {\bibfnamefont {L.}~\bibnamefont {Wang}}, \bibinfo {author} {\bibfnamefont
  {S.}~\bibnamefont {Sorgenfrei}}, \bibinfo {author} {\bibfnamefont
  {K.}~\bibnamefont {Watanabe}}, \bibinfo {author} {\bibfnamefont
  {T.}~\bibnamefont {Taniguchi}}, \bibinfo {author} {\bibfnamefont
  {P.}~\bibnamefont {Kim}}, \bibinfo {author} {\bibfnamefont {K.~L.}\
  \bibnamefont {Shepard}},\ and\ \bibinfo {author} {\bibfnamefont
  {J.}~\bibnamefont {Hone}},\ }\href@noop {} {\bibfield  {journal} {\bibinfo
  {journal} {Nature nanotechnology}\ }\textbf {\bibinfo {volume} {5}},\
  \bibinfo {pages} {722} (\bibinfo {year} {2010})}\BibitemShut {NoStop}%
\bibitem [{\citenamefont {Kretinin}\ \emph {et~al.}(2014)\citenamefont
  {Kretinin}, \citenamefont {Cao}, \citenamefont {Tu}, \citenamefont {Yu},
  \citenamefont {Jalil}, \citenamefont {Novoselov}, \citenamefont {Haigh},
  \citenamefont {Gholinia}, \citenamefont {Mishchenko}, \citenamefont {Lozada},
  \citenamefont {Georgiu}, \citenamefont {Woods}, \citenamefont {Withers},
  \citenamefont {Blake}, \citenamefont {Eda}, \citenamefont {Wirsig},
  \citenamefont {Hucho}, \citenamefont {Watanabe}, \citenamefont {Taniguchi},
  \citenamefont {Geim},\ and\ \citenamefont
  {Gorbachev}}]{kretinin2014electronic}%
  \BibitemOpen
  \bibfield  {author} {\bibinfo {author} {\bibfnamefont {A.}~\bibnamefont
  {Kretinin}}, \bibinfo {author} {\bibfnamefont {Y.}~\bibnamefont {Cao}},
  \bibinfo {author} {\bibfnamefont {J.}~\bibnamefont {Tu}}, \bibinfo {author}
  {\bibfnamefont {G.}~\bibnamefont {Yu}}, \bibinfo {author} {\bibfnamefont
  {R.}~\bibnamefont {Jalil}}, \bibinfo {author} {\bibfnamefont
  {K.}~\bibnamefont {Novoselov}}, \bibinfo {author} {\bibfnamefont
  {S.}~\bibnamefont {Haigh}}, \bibinfo {author} {\bibfnamefont
  {A.}~\bibnamefont {Gholinia}}, \bibinfo {author} {\bibfnamefont
  {A.}~\bibnamefont {Mishchenko}}, \bibinfo {author} {\bibfnamefont
  {M.}~\bibnamefont {Lozada}}, \bibinfo {author} {\bibfnamefont
  {T.}~\bibnamefont {Georgiu}}, \bibinfo {author} {\bibfnamefont
  {C.}~\bibnamefont {Woods}}, \bibinfo {author} {\bibfnamefont
  {F.}~\bibnamefont {Withers}}, \bibinfo {author} {\bibfnamefont
  {P.}~\bibnamefont {Blake}}, \bibinfo {author} {\bibfnamefont
  {G.}~\bibnamefont {Eda}}, \bibinfo {author} {\bibfnamefont {A.}~\bibnamefont
  {Wirsig}}, \bibinfo {author} {\bibfnamefont {C.}~\bibnamefont {Hucho}},
  \bibinfo {author} {\bibfnamefont {K.}~\bibnamefont {Watanabe}}, \bibinfo
  {author} {\bibfnamefont {T.}~\bibnamefont {Taniguchi}}, \bibinfo {author}
  {\bibfnamefont {A.}~\bibnamefont {Geim}},\ and\ \bibinfo {author}
  {\bibfnamefont {R.}~\bibnamefont {Gorbachev}},\ }\href@noop {} {\bibfield
  {journal} {\bibinfo  {journal} {Nano letters}\ }\textbf {\bibinfo {volume}
  {14}},\ \bibinfo {pages} {3270} (\bibinfo {year} {2014})}\BibitemShut
  {NoStop}%
\bibitem [{\citenamefont {Ashhadi}, \citenamefont {Hadavi},\ and\ \citenamefont
  {Sarri}(2017)}]{ashhadi2017electronic}%
  \BibitemOpen
  \bibfield  {author} {\bibinfo {author} {\bibfnamefont {M.}~\bibnamefont
  {Ashhadi}}, \bibinfo {author} {\bibfnamefont {M.}~\bibnamefont {Hadavi}},\
  and\ \bibinfo {author} {\bibfnamefont {Z.}~\bibnamefont {Sarri}},\
  }\href@noop {} {\bibfield  {journal} {\bibinfo  {journal} {Physica E:
  Low-dimensional Systems and Nanostructures}\ }\textbf {\bibinfo {volume}
  {87}},\ \bibinfo {pages} {312} (\bibinfo {year} {2017})}\BibitemShut
  {NoStop}%
\bibitem [{\citenamefont {Epstein}\ \emph {et~al.}(2020)\citenamefont
  {Epstein}, \citenamefont {Chaves}, \citenamefont {Rhodes}, \citenamefont
  {Frank}, \citenamefont {Watanabe}, \citenamefont {Taniguchi}, \citenamefont
  {Giessen}, \citenamefont {Hone}, \citenamefont {Peres},\ and\ \citenamefont
  {Koppens}}]{epstein2020highly}%
  \BibitemOpen
  \bibfield  {author} {\bibinfo {author} {\bibfnamefont {I.}~\bibnamefont
  {Epstein}}, \bibinfo {author} {\bibfnamefont {A.~J.}\ \bibnamefont {Chaves}},
  \bibinfo {author} {\bibfnamefont {D.~A.}\ \bibnamefont {Rhodes}}, \bibinfo
  {author} {\bibfnamefont {B.}~\bibnamefont {Frank}}, \bibinfo {author}
  {\bibfnamefont {K.}~\bibnamefont {Watanabe}}, \bibinfo {author}
  {\bibfnamefont {T.}~\bibnamefont {Taniguchi}}, \bibinfo {author}
  {\bibfnamefont {H.}~\bibnamefont {Giessen}}, \bibinfo {author} {\bibfnamefont
  {J.~C.}\ \bibnamefont {Hone}}, \bibinfo {author} {\bibfnamefont {N.~M.}\
  \bibnamefont {Peres}},\ and\ \bibinfo {author} {\bibfnamefont {F.~H.}\
  \bibnamefont {Koppens}},\ }\href@noop {} {\bibfield  {journal} {\bibinfo
  {journal} {2D Materials}\ }\textbf {\bibinfo {volume} {7}},\ \bibinfo {pages}
  {035031} (\bibinfo {year} {2020})}\BibitemShut {NoStop}%
\bibitem [{\citenamefont {Cudazzo}, \citenamefont {Tokatly},\ and\
  \citenamefont {Rubio}(2011)}]{Cudazzo2011}%
  \BibitemOpen
  \bibfield  {author} {\bibinfo {author} {\bibfnamefont {P.}~\bibnamefont
  {Cudazzo}}, \bibinfo {author} {\bibfnamefont {I.~V.}\ \bibnamefont
  {Tokatly}},\ and\ \bibinfo {author} {\bibfnamefont {A.}~\bibnamefont
  {Rubio}},\ }\href {https://doi.org/10.1103/PhysRevB.84.085406} {\bibfield
  {journal} {\bibinfo  {journal} {Phys. Rev. B}\ }\textbf {\bibinfo {volume}
  {84}},\ \bibinfo {pages} {085406} (\bibinfo {year} {2011})}\BibitemShut
  {NoStop}%
\bibitem [{\citenamefont {Watanabe}, \citenamefont {Taniguchi},\ and\
  \citenamefont {Kanda}(2004)}]{watanabe2004direct}%
  \BibitemOpen
  \bibfield  {author} {\bibinfo {author} {\bibfnamefont {K.}~\bibnamefont
  {Watanabe}}, \bibinfo {author} {\bibfnamefont {T.}~\bibnamefont
  {Taniguchi}},\ and\ \bibinfo {author} {\bibfnamefont {H.}~\bibnamefont
  {Kanda}},\ }\href@noop {} {\bibfield  {journal} {\bibinfo  {journal} {Nature
  materials}\ }\textbf {\bibinfo {volume} {3}},\ \bibinfo {pages} {404}
  (\bibinfo {year} {2004})}\BibitemShut {NoStop}%
\bibitem [{\citenamefont {Kubota}\ \emph {et~al.}(2007)\citenamefont {Kubota},
  \citenamefont {Watanabe}, \citenamefont {Tsuda},\ and\ \citenamefont
  {Taniguchi}}]{kubota2007deep}%
  \BibitemOpen
  \bibfield  {author} {\bibinfo {author} {\bibfnamefont {Y.}~\bibnamefont
  {Kubota}}, \bibinfo {author} {\bibfnamefont {K.}~\bibnamefont {Watanabe}},
  \bibinfo {author} {\bibfnamefont {O.}~\bibnamefont {Tsuda}},\ and\ \bibinfo
  {author} {\bibfnamefont {T.}~\bibnamefont {Taniguchi}},\ }\href@noop {}
  {\bibfield  {journal} {\bibinfo  {journal} {Science}\ }\textbf {\bibinfo
  {volume} {317}},\ \bibinfo {pages} {932} (\bibinfo {year}
  {2007})}\BibitemShut {NoStop}%
\bibitem [{\citenamefont {Fuchs}\ \emph {et~al.}(2008)\citenamefont {Fuchs},
  \citenamefont {R\"odl}, \citenamefont {Schleife},\ and\ \citenamefont
  {Bechstedt}}]{Fuchs2013}%
  \BibitemOpen
  \bibfield  {author} {\bibinfo {author} {\bibfnamefont {F.}~\bibnamefont
  {Fuchs}}, \bibinfo {author} {\bibfnamefont {C.}~\bibnamefont {R\"odl}},
  \bibinfo {author} {\bibfnamefont {A.}~\bibnamefont {Schleife}},\ and\
  \bibinfo {author} {\bibfnamefont {F.}~\bibnamefont {Bechstedt}},\ }\href
  {https://doi.org/10.1103/PhysRevB.78.085103} {\bibfield  {journal} {\bibinfo
  {journal} {Phys. Rev. B}\ }\textbf {\bibinfo {volume} {78}},\ \bibinfo
  {pages} {085103} (\bibinfo {year} {2008})}\BibitemShut {NoStop}%
\bibitem [{\citenamefont {Komsa}\ and\ \citenamefont
  {Krasheninnikov}(2013)}]{Komsa2013}%
  \BibitemOpen
  \bibfield  {author} {\bibinfo {author} {\bibfnamefont {H.-P.}\ \bibnamefont
  {Komsa}}\ and\ \bibinfo {author} {\bibfnamefont {A.~V.}\ \bibnamefont
  {Krasheninnikov}},\ }\href {https://doi.org/10.1103/PhysRevB.88.085318}
  {\bibfield  {journal} {\bibinfo  {journal} {Phys. Rev. B}\ }\textbf {\bibinfo
  {volume} {88}},\ \bibinfo {pages} {085318} (\bibinfo {year}
  {2013})}\BibitemShut {NoStop}%
\bibitem [{\citenamefont {Galvani}\ \emph {et~al.}(2016)\citenamefont
  {Galvani}, \citenamefont {Paleari}, \citenamefont {Miranda}, \citenamefont
  {Molina-S\'anchez}, \citenamefont {Wirtz}, \citenamefont {Latil},
  \citenamefont {Amara},\ and\ \citenamefont {Ducastelle}}]{Alejandro2016}%
  \BibitemOpen
  \bibfield  {author} {\bibinfo {author} {\bibfnamefont {T.}~\bibnamefont
  {Galvani}}, \bibinfo {author} {\bibfnamefont {F.}~\bibnamefont {Paleari}},
  \bibinfo {author} {\bibfnamefont {H.~P.~C.}\ \bibnamefont {Miranda}},
  \bibinfo {author} {\bibfnamefont {A.}~\bibnamefont {Molina-S\'anchez}},
  \bibinfo {author} {\bibfnamefont {L.}~\bibnamefont {Wirtz}}, \bibinfo
  {author} {\bibfnamefont {S.}~\bibnamefont {Latil}}, \bibinfo {author}
  {\bibfnamefont {H.}~\bibnamefont {Amara}},\ and\ \bibinfo {author}
  {\bibfnamefont {F.~m.~c.}\ \bibnamefont {Ducastelle}},\ }\href
  {https://doi.org/10.1103/PhysRevB.94.125303} {\bibfield  {journal} {\bibinfo
  {journal} {Phys. Rev. B}\ }\textbf {\bibinfo {volume} {94}},\ \bibinfo
  {pages} {125303} (\bibinfo {year} {2016})}\BibitemShut {NoStop}%
\bibitem [{\citenamefont {Di~Sabatino}, \citenamefont {Berger},\ and\
  \citenamefont {Romaniello}(2020)}]{di2020optical}%
  \BibitemOpen
  \bibfield  {author} {\bibinfo {author} {\bibfnamefont {S.}~\bibnamefont
  {Di~Sabatino}}, \bibinfo {author} {\bibfnamefont {J.}~\bibnamefont
  {Berger}},\ and\ \bibinfo {author} {\bibfnamefont {P.}~\bibnamefont
  {Romaniello}},\ }\href@noop {} {\bibfield  {journal} {\bibinfo  {journal}
  {Faraday Discussions}\ }\textbf {\bibinfo {volume} {224}},\ \bibinfo {pages}
  {467} (\bibinfo {year} {2020})}\BibitemShut {NoStop}%
\bibitem [{\citenamefont {Taghizadeh}\ and\ \citenamefont
  {Pedersen}(2019)}]{Pedersen2019}%
  \BibitemOpen
  \bibfield  {author} {\bibinfo {author} {\bibfnamefont {A.}~\bibnamefont
  {Taghizadeh}}\ and\ \bibinfo {author} {\bibfnamefont {T.~G.}\ \bibnamefont
  {Pedersen}},\ }\href {https://doi.org/10.1103/PhysRevB.99.235433} {\bibfield
  {journal} {\bibinfo  {journal} {Phys. Rev. B}\ }\textbf {\bibinfo {volume}
  {99}},\ \bibinfo {pages} {235433} (\bibinfo {year} {2019})}\BibitemShut
  {NoStop}%
\bibitem [{\citenamefont {Henriques}\ \emph {et~al.}(2020)\citenamefont
  {Henriques}, \citenamefont {Ventura}, \citenamefont {Fernandes},\ and\
  \citenamefont {Peres}}]{henriques2019optical}%
  \BibitemOpen
  \bibfield  {author} {\bibinfo {author} {\bibfnamefont {J.}~\bibnamefont
  {Henriques}}, \bibinfo {author} {\bibfnamefont {G.}~\bibnamefont {Ventura}},
  \bibinfo {author} {\bibfnamefont {C.}~\bibnamefont {Fernandes}},\ and\
  \bibinfo {author} {\bibfnamefont {N.}~\bibnamefont {Peres}},\ }\href@noop {}
  {\bibfield  {journal} {\bibinfo  {journal} {Journal of Physics: Condensed
  Matter}\ }\textbf {\bibinfo {volume} {32}},\ \bibinfo {pages} {025304}
  (\bibinfo {year} {2020})}\BibitemShut {NoStop}%
\bibitem [{\citenamefont {Ditchfield}, \citenamefont {Hehre},\ and\
  \citenamefont {Pople}(1970)}]{ditchfield1970self}%
  \BibitemOpen
  \bibfield  {author} {\bibinfo {author} {\bibfnamefont {R.}~\bibnamefont
  {Ditchfield}}, \bibinfo {author} {\bibfnamefont {W.}~\bibnamefont {Hehre}},\
  and\ \bibinfo {author} {\bibfnamefont {J.}~\bibnamefont {Pople}},\
  }\href@noop {} {\bibfield  {journal} {\bibinfo  {journal} {The Journal of
  Chemical Physics}\ }\textbf {\bibinfo {volume} {52}},\ \bibinfo {pages}
  {5001} (\bibinfo {year} {1970})}\BibitemShut {NoStop}%
\bibitem [{\citenamefont {Zhang}\ \emph {et~al.}(2014)\citenamefont {Zhang},
  \citenamefont {Wang}, \citenamefont {Chan}, \citenamefont {Manolatou},\ and\
  \citenamefont {Rana}}]{Zhang2014}%
  \BibitemOpen
  \bibfield  {author} {\bibinfo {author} {\bibfnamefont {C.}~\bibnamefont
  {Zhang}}, \bibinfo {author} {\bibfnamefont {H.}~\bibnamefont {Wang}},
  \bibinfo {author} {\bibfnamefont {W.}~\bibnamefont {Chan}}, \bibinfo {author}
  {\bibfnamefont {C.}~\bibnamefont {Manolatou}},\ and\ \bibinfo {author}
  {\bibfnamefont {F.}~\bibnamefont {Rana}},\ }\href
  {https://doi.org/10.1103/PhysRevB.89.205436} {\bibfield  {journal} {\bibinfo
  {journal} {Phys. Rev. B}\ }\textbf {\bibinfo {volume} {89}},\ \bibinfo
  {pages} {205436} (\bibinfo {year} {2014})}\BibitemShut {NoStop}%
\bibitem [{\citenamefont {Quintela}\ and\ \citenamefont
  {Peres}(2020)}]{quintela2020colloquium}%
  \BibitemOpen
  \bibfield  {author} {\bibinfo {author} {\bibfnamefont {M.~F.~M.}\
  \bibnamefont {Quintela}}\ and\ \bibinfo {author} {\bibfnamefont {N.~M.}\
  \bibnamefont {Peres}},\ }\href@noop {} {\bibfield  {journal} {\bibinfo
  {journal} {The European Physical Journal B}\ }\textbf {\bibinfo {volume}
  {93}},\ \bibinfo {pages} {1} (\bibinfo {year} {2020})}\BibitemShut {NoStop}%
\bibitem [{\citenamefont {Ribeiro}\ and\ \citenamefont
  {Peres}(2011)}]{Ribeiro2011}%
  \BibitemOpen
  \bibfield  {author} {\bibinfo {author} {\bibfnamefont {R.~M.}\ \bibnamefont
  {Ribeiro}}\ and\ \bibinfo {author} {\bibfnamefont {N.~M.~R.}\ \bibnamefont
  {Peres}},\ }\href {https://doi.org/10.1103/PhysRevB.83.235312} {\bibfield
  {journal} {\bibinfo  {journal} {Phys. Rev. B}\ }\textbf {\bibinfo {volume}
  {83}},\ \bibinfo {pages} {235312} (\bibinfo {year} {2011})}\BibitemShut
  {NoStop}%
\bibitem [{\citenamefont {Fujimoto}\ and\ \citenamefont
  {Saito}(2016)}]{Fujimoto2016}%
  \BibitemOpen
  \bibfield  {author} {\bibinfo {author} {\bibfnamefont {Y.}~\bibnamefont
  {Fujimoto}}\ and\ \bibinfo {author} {\bibfnamefont {S.}~\bibnamefont
  {Saito}},\ }\href {https://doi.org/10.1103/PhysRevB.94.245427} {\bibfield
  {journal} {\bibinfo  {journal} {Phys. Rev. B}\ }\textbf {\bibinfo {volume}
  {94}},\ \bibinfo {pages} {245427} (\bibinfo {year} {2016})}\BibitemShut
  {NoStop}%
\bibitem [{\citenamefont {Wickramaratne}, \citenamefont {Weston},\ and\
  \citenamefont {Van~de Walle}(2018)}]{wickramaratne2018monolayer}%
  \BibitemOpen
  \bibfield  {author} {\bibinfo {author} {\bibfnamefont {D.}~\bibnamefont
  {Wickramaratne}}, \bibinfo {author} {\bibfnamefont {L.}~\bibnamefont
  {Weston}},\ and\ \bibinfo {author} {\bibfnamefont {C.~G.}\ \bibnamefont
  {Van~de Walle}},\ }\href@noop {} {\bibfield  {journal} {\bibinfo  {journal}
  {The Journal of Physical Chemistry C}\ }\textbf {\bibinfo {volume} {122}},\
  \bibinfo {pages} {25524} (\bibinfo {year} {2018})}\BibitemShut {NoStop}%
\bibitem [{\citenamefont {Lopes~dos Santos}, \citenamefont {Peres},\ and\
  \citenamefont {Castro~Neto}(2007)}]{JLS2007}%
  \BibitemOpen
  \bibfield  {author} {\bibinfo {author} {\bibfnamefont {J.~M.~B.}\
  \bibnamefont {Lopes~dos Santos}}, \bibinfo {author} {\bibfnamefont
  {N.~M.~R.}\ \bibnamefont {Peres}},\ and\ \bibinfo {author} {\bibfnamefont
  {A.~H.}\ \bibnamefont {Castro~Neto}},\ }\href
  {https://doi.org/10.1103/PhysRevLett.99.256802} {\bibfield  {journal}
  {\bibinfo  {journal} {Phys. Rev. Lett.}\ }\textbf {\bibinfo {volume} {99}},\
  \bibinfo {pages} {256802} (\bibinfo {year} {2007})}\BibitemShut {NoStop}%
\bibitem [{\citenamefont {Lopes~dos Santos}, \citenamefont {Peres},\ and\
  \citenamefont {Castro~Neto}(2012)}]{JLS2012}%
  \BibitemOpen
  \bibfield  {author} {\bibinfo {author} {\bibfnamefont {J.~M.~B.}\
  \bibnamefont {Lopes~dos Santos}}, \bibinfo {author} {\bibfnamefont
  {N.~M.~R.}\ \bibnamefont {Peres}},\ and\ \bibinfo {author} {\bibfnamefont
  {A.~H.}\ \bibnamefont {Castro~Neto}},\ }\href
  {https://doi.org/10.1103/PhysRevB.86.155449} {\bibfield  {journal} {\bibinfo
  {journal} {Phys. Rev. B}\ }\textbf {\bibinfo {volume} {86}},\ \bibinfo
  {pages} {155449} (\bibinfo {year} {2012})}\BibitemShut {NoStop}%
\bibitem [{\citenamefont {Xian}\ \emph {et~al.}(2019)\citenamefont {Xian},
  \citenamefont {Kennes}, \citenamefont {Tancogne-Dejean}, \citenamefont
  {Altarelli},\ and\ \citenamefont {Rubio}}]{xian2019multiflat}%
  \BibitemOpen
  \bibfield  {author} {\bibinfo {author} {\bibfnamefont {L.}~\bibnamefont
  {Xian}}, \bibinfo {author} {\bibfnamefont {D.~M.}\ \bibnamefont {Kennes}},
  \bibinfo {author} {\bibfnamefont {N.}~\bibnamefont {Tancogne-Dejean}},
  \bibinfo {author} {\bibfnamefont {M.}~\bibnamefont {Altarelli}},\ and\
  \bibinfo {author} {\bibfnamefont {A.}~\bibnamefont {Rubio}},\ }\href@noop {}
  {\bibfield  {journal} {\bibinfo  {journal} {Nano letters}\ }\textbf {\bibinfo
  {volume} {19}},\ \bibinfo {pages} {4934} (\bibinfo {year}
  {2019})}\BibitemShut {NoStop}%
\bibitem [{\citenamefont {Walet}\ and\ \citenamefont
  {Guinea}(2021)}]{Guinea2021}%
  \BibitemOpen
  \bibfield  {author} {\bibinfo {author} {\bibfnamefont {N.~R.}\ \bibnamefont
  {Walet}}\ and\ \bibinfo {author} {\bibfnamefont {F.}~\bibnamefont {Guinea}},\
  }\href {https://doi.org/10.1103/PhysRevB.103.125427} {\bibfield  {journal}
  {\bibinfo  {journal} {Phys. Rev. B}\ }\textbf {\bibinfo {volume} {103}},\
  \bibinfo {pages} {125427} (\bibinfo {year} {2021})}\BibitemShut {NoStop}%
\bibitem [{\citenamefont {Yao}\ \emph {et~al.}(2021)\citenamefont {Yao},
  \citenamefont {Finney}, \citenamefont {Zhang}, \citenamefont {Moore},
  \citenamefont {Xian}, \citenamefont {Tancogne-Dejean}, \citenamefont {Liu},
  \citenamefont {Ardelean}, \citenamefont {Xu}, \citenamefont {Halbertal} \emph
  {et~al.}}]{yao2021enhanced}%
  \BibitemOpen
  \bibfield  {author} {\bibinfo {author} {\bibfnamefont {K.}~\bibnamefont
  {Yao}}, \bibinfo {author} {\bibfnamefont {N.~R.}\ \bibnamefont {Finney}},
  \bibinfo {author} {\bibfnamefont {J.}~\bibnamefont {Zhang}}, \bibinfo
  {author} {\bibfnamefont {S.~L.}\ \bibnamefont {Moore}}, \bibinfo {author}
  {\bibfnamefont {L.}~\bibnamefont {Xian}}, \bibinfo {author} {\bibfnamefont
  {N.}~\bibnamefont {Tancogne-Dejean}}, \bibinfo {author} {\bibfnamefont
  {F.}~\bibnamefont {Liu}}, \bibinfo {author} {\bibfnamefont {J.}~\bibnamefont
  {Ardelean}}, \bibinfo {author} {\bibfnamefont {X.}~\bibnamefont {Xu}},
  \bibinfo {author} {\bibfnamefont {D.}~\bibnamefont {Halbertal}}, \emph
  {et~al.},\ }\href@noop {} {\bibfield  {journal} {\bibinfo  {journal} {Science
  Advances}\ }\textbf {\bibinfo {volume} {7}},\ \bibinfo {pages} {eabe8691}
  (\bibinfo {year} {2021})}\BibitemShut {NoStop}%
\bibitem [{\citenamefont {Aggoune}\ \emph {et~al.}(2018)\citenamefont
  {Aggoune}, \citenamefont {Cocchi}, \citenamefont {Nabok}, \citenamefont
  {Rezouali}, \citenamefont {Belkhir},\ and\ \citenamefont
  {Draxl}}]{Draxl2018}%
  \BibitemOpen
  \bibfield  {author} {\bibinfo {author} {\bibfnamefont {W.}~\bibnamefont
  {Aggoune}}, \bibinfo {author} {\bibfnamefont {C.}~\bibnamefont {Cocchi}},
  \bibinfo {author} {\bibfnamefont {D.}~\bibnamefont {Nabok}}, \bibinfo
  {author} {\bibfnamefont {K.}~\bibnamefont {Rezouali}}, \bibinfo {author}
  {\bibfnamefont {M.~A.}\ \bibnamefont {Belkhir}},\ and\ \bibinfo {author}
  {\bibfnamefont {C.}~\bibnamefont {Draxl}},\ }\href
  {https://doi.org/10.1103/PhysRevB.97.241114} {\bibfield  {journal} {\bibinfo
  {journal} {Phys. Rev. B}\ }\textbf {\bibinfo {volume} {97}},\ \bibinfo
  {pages} {241114} (\bibinfo {year} {2018})}\BibitemShut {NoStop}%
\bibitem [{\citenamefont {Attaccalite}\ \emph {et~al.}(2018)\citenamefont
  {Attaccalite}, \citenamefont {Gr\"uning}, \citenamefont {Amara},
  \citenamefont {Latil},\ and\ \citenamefont {Ducastelle}}]{Ducastelle2018}%
  \BibitemOpen
  \bibfield  {author} {\bibinfo {author} {\bibfnamefont {C.}~\bibnamefont
  {Attaccalite}}, \bibinfo {author} {\bibfnamefont {M.}~\bibnamefont
  {Gr\"uning}}, \bibinfo {author} {\bibfnamefont {H.}~\bibnamefont {Amara}},
  \bibinfo {author} {\bibfnamefont {S.}~\bibnamefont {Latil}},\ and\ \bibinfo
  {author} {\bibfnamefont {F.~m.~c.}\ \bibnamefont {Ducastelle}},\ }\href
  {https://doi.org/10.1103/PhysRevB.98.165126} {\bibfield  {journal} {\bibinfo
  {journal} {Phys. Rev. B}\ }\textbf {\bibinfo {volume} {98}},\ \bibinfo
  {pages} {165126} (\bibinfo {year} {2018})}\BibitemShut {NoStop}%
\bibitem [{\citenamefont {Paleari}\ \emph {et~al.}(2018)\citenamefont
  {Paleari}, \citenamefont {Galvani}, \citenamefont {Amara}, \citenamefont
  {Ducastelle}, \citenamefont {Molina-S{\'a}nchez},\ and\ \citenamefont
  {Wirtz}}]{paleari2018excitons}%
  \BibitemOpen
  \bibfield  {author} {\bibinfo {author} {\bibfnamefont {F.}~\bibnamefont
  {Paleari}}, \bibinfo {author} {\bibfnamefont {T.}~\bibnamefont {Galvani}},
  \bibinfo {author} {\bibfnamefont {H.}~\bibnamefont {Amara}}, \bibinfo
  {author} {\bibfnamefont {F.}~\bibnamefont {Ducastelle}}, \bibinfo {author}
  {\bibfnamefont {A.}~\bibnamefont {Molina-S{\'a}nchez}},\ and\ \bibinfo
  {author} {\bibfnamefont {L.}~\bibnamefont {Wirtz}},\ }\href@noop {}
  {\bibfield  {journal} {\bibinfo  {journal} {2D Materials}\ }\textbf {\bibinfo
  {volume} {5}},\ \bibinfo {pages} {045017} (\bibinfo {year}
  {2018})}\BibitemShut {NoStop}%
\bibitem [{\citenamefont {Mengle}\ and\ \citenamefont
  {Kioupakis}(2019)}]{mengle2019impact}%
  \BibitemOpen
  \bibfield  {author} {\bibinfo {author} {\bibfnamefont {K.}~\bibnamefont
  {Mengle}}\ and\ \bibinfo {author} {\bibfnamefont {E.}~\bibnamefont
  {Kioupakis}},\ }\href@noop {} {\bibfield  {journal} {\bibinfo  {journal} {APL
  Materials}\ }\textbf {\bibinfo {volume} {7}},\ \bibinfo {pages} {021106}
  (\bibinfo {year} {2019})}\BibitemShut {NoStop}%
\bibitem [{\citenamefont {Suzuki}\ and\ \citenamefont
  {Watanabe}(2020)}]{suzuki2020excitons}%
  \BibitemOpen
  \bibfield  {author} {\bibinfo {author} {\bibfnamefont {Y.}~\bibnamefont
  {Suzuki}}\ and\ \bibinfo {author} {\bibfnamefont {K.}~\bibnamefont
  {Watanabe}},\ }\href@noop {} {\bibfield  {journal} {\bibinfo  {journal}
  {Physical Chemistry Chemical Physics}\ }\textbf {\bibinfo {volume} {22}},\
  \bibinfo {pages} {2908} (\bibinfo {year} {2020})}\BibitemShut {NoStop}%
\bibitem [{\citenamefont {Paradisanos}\ \emph {et~al.}(2020)\citenamefont
  {Paradisanos}, \citenamefont {Shree}, \citenamefont {George}, \citenamefont
  {Leisgang}, \citenamefont {Robert}, \citenamefont {Watanabe}, \citenamefont
  {Taniguchi}, \citenamefont {Warburton}, \citenamefont {Turchanin},
  \citenamefont {Marie}, \citenamefont {Gerber},\ and\ \citenamefont
  {Urbaszek}}]{paradisanos2020controlling}%
  \BibitemOpen
  \bibfield  {author} {\bibinfo {author} {\bibfnamefont {I.}~\bibnamefont
  {Paradisanos}}, \bibinfo {author} {\bibfnamefont {S.}~\bibnamefont {Shree}},
  \bibinfo {author} {\bibfnamefont {A.}~\bibnamefont {George}}, \bibinfo
  {author} {\bibfnamefont {N.}~\bibnamefont {Leisgang}}, \bibinfo {author}
  {\bibfnamefont {C.}~\bibnamefont {Robert}}, \bibinfo {author} {\bibfnamefont
  {K.}~\bibnamefont {Watanabe}}, \bibinfo {author} {\bibfnamefont
  {T.}~\bibnamefont {Taniguchi}}, \bibinfo {author} {\bibfnamefont {R.~J.}\
  \bibnamefont {Warburton}}, \bibinfo {author} {\bibfnamefont {A.}~\bibnamefont
  {Turchanin}}, \bibinfo {author} {\bibfnamefont {X.}~\bibnamefont {Marie}},
  \bibinfo {author} {\bibfnamefont {I.~C.}\ \bibnamefont {Gerber}},\ and\
  \bibinfo {author} {\bibfnamefont {B.}~\bibnamefont {Urbaszek}},\ }\href@noop
  {} {\bibfield  {journal} {\bibinfo  {journal} {Nature communications}\
  }\textbf {\bibinfo {volume} {11}},\ \bibinfo {pages} {1} (\bibinfo {year}
  {2020})}\BibitemShut {NoStop}%
\bibitem [{\citenamefont {Rytova}(1967)}]{rytova1967}%
  \BibitemOpen
  \bibfield  {author} {\bibinfo {author} {\bibfnamefont {S.}~\bibnamefont
  {Rytova}},\ }\href@noop {} {\bibfield  {journal} {\bibinfo  {journal} {Moscow
  University Physics Bulletin}\ }\textbf {\bibinfo {volume} {22}} (\bibinfo
  {year} {1967})}\BibitemShut {NoStop}%
\bibitem [{\citenamefont {Keldysh}(1979)}]{keldysh1979coulomb}%
  \BibitemOpen
  \bibfield  {author} {\bibinfo {author} {\bibfnamefont {L.}~\bibnamefont
  {Keldysh}},\ }\href@noop {} {\bibfield  {journal} {\bibinfo  {journal} {Sov.
  J. Exp. and Theor. Phys. Lett.}\ }\textbf {\bibinfo {volume} {29}},\ \bibinfo
  {pages} {658} (\bibinfo {year} {1979})}\BibitemShut {NoStop}%
\bibitem [{\citenamefont {Tian}\ \emph {et~al.}(2019)\citenamefont {Tian},
  \citenamefont {Scullion}, \citenamefont {Hughes}, \citenamefont {Li},
  \citenamefont {Shih}, \citenamefont {Coleman}, \citenamefont {Chhowalla},\
  and\ \citenamefont {Santos}}]{tian2019electronic}%
  \BibitemOpen
  \bibfield  {author} {\bibinfo {author} {\bibfnamefont {T.}~\bibnamefont
  {Tian}}, \bibinfo {author} {\bibfnamefont {D.}~\bibnamefont {Scullion}},
  \bibinfo {author} {\bibfnamefont {D.}~\bibnamefont {Hughes}}, \bibinfo
  {author} {\bibfnamefont {L.~H.}\ \bibnamefont {Li}}, \bibinfo {author}
  {\bibfnamefont {C.-J.}\ \bibnamefont {Shih}}, \bibinfo {author}
  {\bibfnamefont {J.}~\bibnamefont {Coleman}}, \bibinfo {author} {\bibfnamefont
  {M.}~\bibnamefont {Chhowalla}},\ and\ \bibinfo {author} {\bibfnamefont
  {E.~J.}\ \bibnamefont {Santos}},\ }\href@noop {} {\bibfield  {journal}
  {\bibinfo  {journal} {Nano letters}\ }\textbf {\bibinfo {volume} {20}},\
  \bibinfo {pages} {841} (\bibinfo {year} {2019})}\BibitemShut {NoStop}%
\bibitem [{\citenamefont {Park}\ and\ \citenamefont
  {Louie}(2010)}]{park2010tunable}%
  \BibitemOpen
  \bibfield  {author} {\bibinfo {author} {\bibfnamefont {C.-H.}\ \bibnamefont
  {Park}}\ and\ \bibinfo {author} {\bibfnamefont {S.~G.}\ \bibnamefont
  {Louie}},\ }\href@noop {} {\bibfield  {journal} {\bibinfo  {journal} {Nano
  letters}\ }\textbf {\bibinfo {volume} {10}},\ \bibinfo {pages} {426}
  (\bibinfo {year} {2010})}\BibitemShut {NoStop}%
\bibitem [{\citenamefont {Cao}, \citenamefont {Wu},\ and\ \citenamefont
  {Louie}(2018)}]{Cao2018}%
  \BibitemOpen
  \bibfield  {author} {\bibinfo {author} {\bibfnamefont {T.}~\bibnamefont
  {Cao}}, \bibinfo {author} {\bibfnamefont {M.}~\bibnamefont {Wu}},\ and\
  \bibinfo {author} {\bibfnamefont {S.~G.}\ \bibnamefont {Louie}},\ }\href
  {https://doi.org/10.1103/PhysRevLett.120.087402} {\bibfield  {journal}
  {\bibinfo  {journal} {Phys. Rev. Lett.}\ }\textbf {\bibinfo {volume} {120}},\
  \bibinfo {pages} {087402} (\bibinfo {year} {2018})}\BibitemShut {NoStop}%
\bibitem [{\citenamefont {Chao}\ and\ \citenamefont {Chuang}(1991)}]{Chao1991}%
  \BibitemOpen
  \bibfield  {author} {\bibinfo {author} {\bibfnamefont {C.~Y.-P.}\
  \bibnamefont {Chao}}\ and\ \bibinfo {author} {\bibfnamefont {S.~L.}\
  \bibnamefont {Chuang}},\ }\href {https://doi.org/10.1103/PhysRevB.43.6530}
  {\bibfield  {journal} {\bibinfo  {journal} {Phys. Rev. B}\ }\textbf {\bibinfo
  {volume} {43}},\ \bibinfo {pages} {6530} (\bibinfo {year}
  {1991})}\BibitemShut {NoStop}%
\bibitem [{\citenamefont {Pedersen}(2015)}]{Thomas2015cond}%
  \BibitemOpen
  \bibfield  {author} {\bibinfo {author} {\bibfnamefont {T.~G.}\ \bibnamefont
  {Pedersen}},\ }\href {https://doi.org/10.1103/PhysRevB.92.235432} {\bibfield
  {journal} {\bibinfo  {journal} {Phys. Rev. B}\ }\textbf {\bibinfo {volume}
  {92}},\ \bibinfo {pages} {235432} (\bibinfo {year} {2015})}\BibitemShut
  {NoStop}%
\bibitem [{\citenamefont {Zhang}, \citenamefont {Shan},\ and\ \citenamefont
  {Xiao}(2018)}]{Zhang2018}%
  \BibitemOpen
  \bibfield  {author} {\bibinfo {author} {\bibfnamefont {X.}~\bibnamefont
  {Zhang}}, \bibinfo {author} {\bibfnamefont {W.-Y.}\ \bibnamefont {Shan}},\
  and\ \bibinfo {author} {\bibfnamefont {D.}~\bibnamefont {Xiao}},\ }\href
  {https://doi.org/10.1103/PhysRevLett.120.077401} {\bibfield  {journal}
  {\bibinfo  {journal} {Phys. Rev. Lett.}\ }\textbf {\bibinfo {volume} {120}},\
  \bibinfo {pages} {077401} (\bibinfo {year} {2018})}\BibitemShut {NoStop}%
\bibitem [{\citenamefont {L{\"o}wdin}(1951)}]{lowdin1951note}%
  \BibitemOpen
  \bibfield  {author} {\bibinfo {author} {\bibfnamefont {P.-O.}\ \bibnamefont
  {L{\"o}wdin}},\ }\href@noop {} {\bibfield  {journal} {\bibinfo  {journal}
  {The Journal of Chemical Physics}\ }\textbf {\bibinfo {volume} {19}},\
  \bibinfo {pages} {1396} (\bibinfo {year} {1951})}\BibitemShut {NoStop}%
\bibitem [{\citenamefont {Winkler}\ \emph {et~al.}(2003)\citenamefont
  {Winkler}, \citenamefont {Papadakis}, \citenamefont {De~Poortere},\ and\
  \citenamefont {Shayegan}}]{winkler2003spin}%
  \BibitemOpen
  \bibfield  {author} {\bibinfo {author} {\bibfnamefont {R.}~\bibnamefont
  {Winkler}}, \bibinfo {author} {\bibfnamefont {S.}~\bibnamefont {Papadakis}},
  \bibinfo {author} {\bibfnamefont {E.}~\bibnamefont {De~Poortere}},\ and\
  \bibinfo {author} {\bibfnamefont {M.}~\bibnamefont {Shayegan}},\ }\href@noop
  {} {\emph {\bibinfo {title} {Spin-Orbit Coupling in Two-Dimensional Electron
  and Hole Systems}}},\ Vol.~\bibinfo {volume} {41}\ (\bibinfo  {publisher}
  {Springer},\ \bibinfo {year} {2003})\BibitemShut {NoStop}%
\bibitem [{\citenamefont {Giannozzi}\ \emph {et~al.}(2009)\citenamefont
  {Giannozzi}, \citenamefont {Baroni}, \citenamefont {Bonini}, \citenamefont
  {Calandra}, \citenamefont {Car}, \citenamefont {Cavazzoni}, \citenamefont
  {Ceresoli}, \citenamefont {Chiarotti}, \citenamefont {Cococcioni},
  \citenamefont {Dabo}, \citenamefont {{Dal Corso}}, \citenamefont
  {de~Gironcoli}, \citenamefont {Fabris}, \citenamefont {Fratesi},
  \citenamefont {Gebauer}, \citenamefont {Gerstmann}, \citenamefont
  {Gougoussis}, \citenamefont {Kokalj}, \citenamefont {Lazzeri}, \citenamefont
  {Martin-Samos}, \citenamefont {Marzari}, \citenamefont {Mauri}, \citenamefont
  {Mazzarello}, \citenamefont {Paolini}, \citenamefont {Pasquarello},
  \citenamefont {Paulatto}, \citenamefont {Sbraccia}, \citenamefont {Scandolo},
  \citenamefont {Sclauzero}, \citenamefont {Seitsonen}, \citenamefont
  {Smogunov}, \citenamefont {Umari},\ and\ \citenamefont
  {Wentzcovitch}}]{QE-2009}%
  \BibitemOpen
  \bibfield  {author} {\bibinfo {author} {\bibfnamefont {P.}~\bibnamefont
  {Giannozzi}}, \bibinfo {author} {\bibfnamefont {S.}~\bibnamefont {Baroni}},
  \bibinfo {author} {\bibfnamefont {N.}~\bibnamefont {Bonini}}, \bibinfo
  {author} {\bibfnamefont {M.}~\bibnamefont {Calandra}}, \bibinfo {author}
  {\bibfnamefont {R.}~\bibnamefont {Car}}, \bibinfo {author} {\bibfnamefont
  {C.}~\bibnamefont {Cavazzoni}}, \bibinfo {author} {\bibfnamefont
  {D.}~\bibnamefont {Ceresoli}}, \bibinfo {author} {\bibfnamefont {G.~L.}\
  \bibnamefont {Chiarotti}}, \bibinfo {author} {\bibfnamefont {M.}~\bibnamefont
  {Cococcioni}}, \bibinfo {author} {\bibfnamefont {I.}~\bibnamefont {Dabo}},
  \bibinfo {author} {\bibfnamefont {A.}~\bibnamefont {{Dal Corso}}}, \bibinfo
  {author} {\bibfnamefont {S.}~\bibnamefont {de~Gironcoli}}, \bibinfo {author}
  {\bibfnamefont {S.}~\bibnamefont {Fabris}}, \bibinfo {author} {\bibfnamefont
  {G.}~\bibnamefont {Fratesi}}, \bibinfo {author} {\bibfnamefont
  {R.}~\bibnamefont {Gebauer}}, \bibinfo {author} {\bibfnamefont
  {U.}~\bibnamefont {Gerstmann}}, \bibinfo {author} {\bibfnamefont
  {C.}~\bibnamefont {Gougoussis}}, \bibinfo {author} {\bibfnamefont
  {A.}~\bibnamefont {Kokalj}}, \bibinfo {author} {\bibfnamefont
  {M.}~\bibnamefont {Lazzeri}}, \bibinfo {author} {\bibfnamefont
  {L.}~\bibnamefont {Martin-Samos}}, \bibinfo {author} {\bibfnamefont
  {N.}~\bibnamefont {Marzari}}, \bibinfo {author} {\bibfnamefont
  {F.}~\bibnamefont {Mauri}}, \bibinfo {author} {\bibfnamefont
  {R.}~\bibnamefont {Mazzarello}}, \bibinfo {author} {\bibfnamefont
  {S.}~\bibnamefont {Paolini}}, \bibinfo {author} {\bibfnamefont
  {A.}~\bibnamefont {Pasquarello}}, \bibinfo {author} {\bibfnamefont
  {L.}~\bibnamefont {Paulatto}}, \bibinfo {author} {\bibfnamefont
  {C.}~\bibnamefont {Sbraccia}}, \bibinfo {author} {\bibfnamefont
  {S.}~\bibnamefont {Scandolo}}, \bibinfo {author} {\bibfnamefont
  {G.}~\bibnamefont {Sclauzero}}, \bibinfo {author} {\bibfnamefont {A.~P.}\
  \bibnamefont {Seitsonen}}, \bibinfo {author} {\bibfnamefont {A.}~\bibnamefont
  {Smogunov}}, \bibinfo {author} {\bibfnamefont {P.}~\bibnamefont {Umari}},\
  and\ \bibinfo {author} {\bibfnamefont {R.~M.}\ \bibnamefont {Wentzcovitch}},\
  }\href {http://www.quantum-espresso.org} {\bibfield  {journal} {\bibinfo
  {journal} {Journal of Physics: Condensed Matter}\ }\textbf {\bibinfo {volume}
  {21}},\ \bibinfo {pages} {395502 (19pp)} (\bibinfo {year}
  {2009})}\BibitemShut {NoStop}%
\bibitem [{\citenamefont {Hamann}(2013)}]{Hamman2013}%
  \BibitemOpen
  \bibfield  {author} {\bibinfo {author} {\bibfnamefont {D.~R.}\ \bibnamefont
  {Hamann}},\ }\href {https://doi.org/10.1103/PhysRevB.88.085117} {\bibfield
  {journal} {\bibinfo  {journal} {Phys. Rev. B}\ }\textbf {\bibinfo {volume}
  {88}},\ \bibinfo {pages} {085117} (\bibinfo {year} {2013})}\BibitemShut
  {NoStop}%
\bibitem [{\citenamefont {Schlipf}\ and\ \citenamefont
  {Gygi}(2015)}]{schlipf2015optimization}%
  \BibitemOpen
  \bibfield  {author} {\bibinfo {author} {\bibfnamefont {M.}~\bibnamefont
  {Schlipf}}\ and\ \bibinfo {author} {\bibfnamefont {F.}~\bibnamefont {Gygi}},\
  }\href@noop {} {\bibfield  {journal} {\bibinfo  {journal} {Computer Physics
  Communications}\ }\textbf {\bibinfo {volume} {196}},\ \bibinfo {pages} {36}
  (\bibinfo {year} {2015})}\BibitemShut {NoStop}%
\bibitem [{\citenamefont {Perdew}, \citenamefont {Burke},\ and\ \citenamefont
  {Ernzerhof}(1996)}]{PhysRevLett.77.3865}%
  \BibitemOpen
  \bibfield  {author} {\bibinfo {author} {\bibfnamefont {J.~P.}\ \bibnamefont
  {Perdew}}, \bibinfo {author} {\bibfnamefont {K.}~\bibnamefont {Burke}},\ and\
  \bibinfo {author} {\bibfnamefont {M.}~\bibnamefont {Ernzerhof}},\ }\href
  {https://doi.org/10.1103/PhysRevLett.77.3865} {\bibfield  {journal} {\bibinfo
   {journal} {Phys. Rev. Lett.}\ }\textbf {\bibinfo {volume} {77}},\ \bibinfo
  {pages} {3865} (\bibinfo {year} {1996})}\BibitemShut {NoStop}%
\bibitem [{\citenamefont {Monkhorst}\ and\ \citenamefont
  {Pack}(1976)}]{PhysRevB.13.5188}%
  \BibitemOpen
  \bibfield  {author} {\bibinfo {author} {\bibfnamefont {H.~J.}\ \bibnamefont
  {Monkhorst}}\ and\ \bibinfo {author} {\bibfnamefont {J.~D.}\ \bibnamefont
  {Pack}},\ }\href {https://doi.org/10.1103/PhysRevB.13.5188} {\bibfield
  {journal} {\bibinfo  {journal} {Phys. Rev. B}\ }\textbf {\bibinfo {volume}
  {13}},\ \bibinfo {pages} {5188} (\bibinfo {year} {1976})}\BibitemShut
  {NoStop}%
\bibitem [{\citenamefont {Grimme}(2006)}]{grimme2006semiempirical}%
  \BibitemOpen
  \bibfield  {author} {\bibinfo {author} {\bibfnamefont {S.}~\bibnamefont
  {Grimme}},\ }\href@noop {} {\bibfield  {journal} {\bibinfo  {journal}
  {Journal of computational chemistry}\ }\textbf {\bibinfo {volume} {27}},\
  \bibinfo {pages} {1787} (\bibinfo {year} {2006})}\BibitemShut {NoStop}%
\bibitem [{\citenamefont {Barone}\ \emph {et~al.}(2009)\citenamefont {Barone},
  \citenamefont {Casarin}, \citenamefont {Forrer}, \citenamefont {Pavone},
  \citenamefont {Sambi},\ and\ \citenamefont {Vittadini}}]{barone2009role}%
  \BibitemOpen
  \bibfield  {author} {\bibinfo {author} {\bibfnamefont {V.}~\bibnamefont
  {Barone}}, \bibinfo {author} {\bibfnamefont {M.}~\bibnamefont {Casarin}},
  \bibinfo {author} {\bibfnamefont {D.}~\bibnamefont {Forrer}}, \bibinfo
  {author} {\bibfnamefont {M.}~\bibnamefont {Pavone}}, \bibinfo {author}
  {\bibfnamefont {M.}~\bibnamefont {Sambi}},\ and\ \bibinfo {author}
  {\bibfnamefont {A.}~\bibnamefont {Vittadini}},\ }\href@noop {} {\bibfield
  {journal} {\bibinfo  {journal} {Journal of computational chemistry}\ }\textbf
  {\bibinfo {volume} {30}},\ \bibinfo {pages} {934} (\bibinfo {year}
  {2009})}\BibitemShut {NoStop}%
\bibitem [{\citenamefont {Kythe}\ and\ \citenamefont
  {Puri}(2011)}]{kythe2011computational}%
  \BibitemOpen
  \bibfield  {author} {\bibinfo {author} {\bibfnamefont {P.}~\bibnamefont
  {Kythe}}\ and\ \bibinfo {author} {\bibfnamefont {P.}~\bibnamefont {Puri}},\
  }\href@noop {} {\emph {\bibinfo {title} {Computational methods for linear
  integral equations}}}\ (\bibinfo  {publisher} {Springer Science \& Business
  Media},\ \bibinfo {year} {2011})\BibitemShut {NoStop}%
\end{thebibliography}

%

\end{document}